\documentclass[fleqn,usenatbib]{mnras}

\usepackage{newtxtext,newtxmath}
\usepackage[version=4]{mhchem}

\usepackage[T1]{fontenc}

\DeclareRobustCommand{\VAN}[3]{#2}
\let\VANthebibliography\thebibliography
\def\thebibliography{\DeclareRobustCommand{\VAN}[3]{##3}\VANthebibliography}

\usepackage{graphicx}	
\usepackage{amsmath}	
\DeclareMathOperator{\EX}{\mathbb{E}}

\title[Interpretable Machine Learning in Astrochemistry]{Understanding Molecular Abundances in Star-Forming Regions Using Interpretable Machine Learning}

\author[J. Heyl et al.]{
Johannes Heyl$^{1}$\thanks{E-mail: johannes.heyl.19@ucl.ac.uk},
Joshua Butterworth$^{2}$,
and Serena Viti$^{2,1}$
\\
$^{1}$Department of Physics and Astronomy, University College London, Gower Street, WC1E 6BT, London, UK \\
$^{2}$Leiden Observatory, Leiden University, Huygens Laboratory, Niels Bohrweg 2, NL-2333 CA Leiden, The Netherlands\\
}

\date{Accepted XXX. Received YYY; in original form ZZZ}

\pubyear{2015}

\begin{document}
\label{firstpage}
\pagerange{\pageref{firstpage}--\pageref{lastpage}}
\maketitle

\begin{abstract}
Astrochemical modelling of the interstellar medium typically makes use of complex computational codes with parameters whose values can be varied. It is not always clear what the exact nature of the relationship is between these input parameters and the output molecular abundances. In this work, a feature importance analysis is conducted using SHapley Additive exPlanations (SHAP), an interpretable machine learning technique, to identify the most important physical parameters as well as their relationship with each output. The outputs are the abundances of species and ratios of abundances. In order to reduce the time taken for this process, a neural network emulator is trained to model each species' output abundance and this emulator is used to perform the interpretable machine learning. SHAP is then used to further explore the relationship between the physical features and the abundances for the various species and ratios we considered. \ce{H2O} and CO's gas phase abundances are found to strongly depend on the metallicity. \ce{NH3} has a strong temperature dependence, with there being two temperature regimes (< 100 K and > 100K). By analysing the chemical network, we relate this to the chemical reactions in our network and find the increased temperature results in increased efficiency of destruction pathways. We investigate the HCN/HNC ratio and show that it can be used as a cosmic thermometer, agreeing with the literature. This ratio is also found to be correlated with the metallicity. The HCN/CS ratio serves as a density tracer, but also has three separate temperature-dependence regimes, which are linked to the chemistry of the two molecules. 

\end{abstract}




\section{Introduction} \label{sec:intro}
Modelling the interstellar medium and star formation is often a complex matter. This is normally done using computational codes that take in a number of physical parameters and use these to integrate the system of coupled ordinary differential equations (ODEs) that represent a chemical network \citep{GRAINOBLE,Nautilus_paper, UCLCHEM}. However, due to the non-linear nature of the chemistry, it is often unclear what the exact relationship is between the initial parameters and the output chemical abundances of the molecules of interest. This is often complicated by the fact that the various parameters have differing effects on the output abundances for different ranges. 

It has been customary in astrochemistry to consider grids of models in which the various parameters are varied \citep{GRAINOBLE, grid_paper, Viti_2017, Cecilia_grids, Tom_paper, Jon_HITS_paper, chemical_modelling}. The time-consuming and computationally expensive nature of many computational codes often limits the total number of model evaluations possible. This makes drawing conclusions about the importance of various parameters difficult. In this work, we look to address both of these issues. We make use of SHapley Additive exPlanations (SHAP) \citep{SHAP_paper} to help improve our understanding of a chemical code. SHAP provides us with a means of understanding why a machine learning model outputs a particular value. By considering various combinations of inputs and outputs, these techniques will tell us what the relationship is. This has found use in astrophysics recently \citep{shaping_the_gas,dust_properties} in the context of interpreting the outputs of machine learning models.

To improve the efficiency of this process, we employ statistical emulation. The process of statistical emulation involves fitting a statistical function to model the relationship between the inputs and outputs of a forward model \citep{emulator_definition}. A significant amount of work has been done in recent years in applying statistical emulation to astrochemistry. \cite{Damien} used a feed-forward neural network to accelerate the Bayesian inference process, while \cite{Grassi_emulator} used these to accelerate the forward modelling. \cite{PINNS_chemistry} considered how a physics-informed neural network could be used to reduce the computational cost of predicting the time evolution of chemistry. \cite{chemulator} utilised autoencoders to model temperature and abundance time evolution.

We adopt the approach taken by \cite{Damien} and \cite{Grassi_emulator} in this work by using an emulator to simulate the final outputs of a chemical code and then evaluate these a number of times for the purposes of the machine learning interpretability algorithm.  Work has been done in this area to  simplify chemical networks to improve interpretability \citep{Hoffmann, Grassi_autoencoders}, but this is not an approach we wish to to consider. Instead we build on the work done in \cite{Damien} and look to use the interpretability techniques on these emulators. The purpose of using an emulator is that it accurately predicts the output of the forward model it is emulating in a fraction of the time. Furthermore, if the emulator is an accurate approximation for the forward model output, then it stands to reason that it accurately captures the mapping between the input parameters and the output. By using machine learning interpretability algorithms, we can identify these. 

In Section \ref{sec:chemical_code}, we introduce the chemical code that we will be looking to emulate. In Section \ref{sec:emulation} we introduce statistical emulation and machine learning interpretability. Section \ref{sec:results} is dedicated to discussing the results of the analysis.

\section{The Chemical Code and Network}\label{sec:chemical_code}
In this work, we use the open source publicly available time-dependent astrochemical code UCLCHEM \citep{UCLCHEM}. This astrochemical code has been developed with several updates \citep{Viti_2004, UCLCHEM_update, UCLCHEM}. UCLCHEM is a time-dependent gas-grain astrochemical code. It utilises a rate equation approach to modelling the abundances of the gas phase and surface species. The initial elemental abundances are listed in Table \ref{elemental_abundances}. The default values in the code for the radiation field and the cosmic ray ionisation rate are $\psi = 1$ Habing and $\zeta = 1.3 \times 10^{-17}$ s$^{-1}$. Radiation is attenuated by the visual extinction. Gas and dust are rescaled from solar values. Extensive documentation on the inner workings of UCLCHEM can be found on the GitHub page\footnote{https://uclchem.github.io/}.

In this work, we use UCLCHEM in two phases of modelling. Phase 1 corresponds to the isothermal gravitational collapse of a diffuse gas cloud modelled as a Bonnor-Ebert sphere. However, this stops once the internal pressure begins to balance out the gravitational pressure. This increase in internal pressure is accompanied by an increase in temperature which is when Phase 2 begins, which models a protostar. At this point, the temperature continues to increase and grain-surface species begin to evaporate as the temperatures near their respective evaporation temperatures.

In Phase 1, the gas cloud collapses isothermally at 10 K from 100 cm$^{-3}$ to some final density, which is left as a free parameter. Phase 2 starts off at this density and begins to heat up. It is Phase 2 that has a number of physical parameters that can be varied in order to model various star-forming scenarios. There are a number of free parameters that we vary in this work, which are the same as in \cite{Damien}. These are: 
\begin{itemize}
    \item Final density of Phase 1 or initial density of Phase 2 (cm$^{-3}$)
    \item Metallicity (a scaling factor of all abundances) 
    \item Radiation Field (Habing)
    \item Cosmic ray ionisation rate (in units of 1.3x10$^{-17}$ s$^{-1}$)
    \item Final Temperature of Phase 2 (in Kelvin)
\end{itemize}

The ranges over which we vary the parameters are summarised in Table \ref{parameter_range_table}.

The grain-surface network we utilise is the default one in the GitHub repository that has been able to reproduce the abundances of the main observed grain-surface species for example in \cite{UCLCHEM} and \cite{chemical_modelling}. The grain-surface reaction mechanisms that are used in UCLCHEM include the Eley-Rideal mechanism as well as the Langmuir-Hinshelwood grain-surface diffusion mechanism. These were implemented into the code in \cite{Quenard}, along with the competition formula from \cite{Chang} and \cite{Garrod_CO2}. The binding energies that are required in order to calculate diffusion reaction rates are taken from \cite{Wakelam}. The gas-phase network is taken from UMIST \citep{UMIST}. While the grain network has undergone minor modifications since \cite{Damien}, the gas network has remained the same. Since we are only considering gas-phase species, minor modifications to the grain network are unlikely to be influential.

\begin{table}
\begin{tabular}{lllll}
\hline
\multicolumn{5}{c}{Parameter ranges}                                                                          \\ \hline
Parameter & Minimum               & Maximum               & Unit   & Scale                                           \\ \hline
n         & 10$^4$ & \textbf{10$^7$} & cm$^{-3}$ &  Logarithmic                           \\
$\zeta$      & 1                     & 1000                  & 1.3x10$^{-17}$ s$^{-1}$ & Logarithmic\\
T         & 10                    & 200                   & K       & Linear                                          \\
m$_z$      & 0                     & 2                     & \textbf{solar value}   & Linear\\
$\psi$      & 1                     & $10^{3}$                     & Habing   & Logarithmic

\\
\hline
\end{tabular}
\caption{The range of values used for each parameter as well as their units and scales. In the context of the machine learning application in this work, we refer to these parameters as the features of the model. } 
\label{parameter_range_table}
\end{table}

\section{Machine Learning Interpretability and Statistical Emulation}\label{sec:emulation}
\subsection{Machine Learning Interpretability}
It is often unclear why a model provides a certain output for a given input. This is not exclusive to machine learning algorithms, but can also be an issue with computational codes that integrate systems of differential equations, such as UCLCHEM. As a result, identifying the effect that a specific physical parameter, which we refer to as a feature in this work, has on the output becomes difficult. The concept of feature importance refers to the size of the contribution of a specific feature in determining the model output. There exist many methods by which one can interpret the effect of a parameter in making a certain prediction value, such as permutation feature importance or Local Surrogate Models. For an overview of the various methods, see \cite{molnar2022}. 

We use Shapley values, a method from game theory,  to quantify the importance of the features \citep{Shapley}. This is the first such application in the area of astrochemistry. While we provide an overview of the method we use in this paper below, we refer the reader to \cite{Shapley}, \cite{SHAP_paper} and \cite{TreeSHAP} for further details.

The Shapley value of the $i$th feature, $\phi_i^j$,\footnote{Unless otherwise specified, superscripts are used throughout this work as labels, not for exponentiation.} is defined as the marginal contribution of that feature in mapping the $j^{th}$ data point in our dataset, $x^j$, to its corresponding output $f(x^j)$ averaged over all possible coalitions. A coalition is defined as a subset of the set of features. Notice that in this case, the function $f$ corresponds to UCLCHEM and $x^j$ corresponds to a particular input vector consisting of one entry for each feature in Table \ref{parameter_range_table} that we modify. Each Shapley value, $\phi_i^j$, is specific to each parameter of each data point.

Shapley value explanations are given as a linear model \citep{molnar2022}. We define a feature explanation model, $\hat{g}$ in the following way: 

\begin{equation}\label{explanation_model_equation}
\hat{g}(x'^j) = \phi_{0} + \sum_{i=1}^{n}\phi_{i}{x'^{j}_{i}},
\end{equation}

where $\phi_{0} = \EX[f(x)]$ is the value of the average prediction in our dataset, $\phi_{i}$ is the explained feature effect of the $i$th feature, $n$ is the number of features and $x'^{j}_{i}$ is an element of the ``coalition vector", $x'^j$, where $x'^j \in \{0, 1\}^{n}$. The coalition vector is a vector consisting of zeros and ones with a zero indicating that a feature is ``absent" and a one indicating it is ``present".  

One can imagine that this feature explanation model gives us an understanding of what happens when we choose to remove certain features, that is set a particular $x'^{j}_{i}$ to equal zero. If we want to be able to calculate the feature importance of a specific feature, then we need to be able to selectively ``remove" features and see how this impacts our model output. When we say that we ``remove" a feature, what we effectively mean is that we replace that value in the input vector by a random value from the dataset for that feature. The logic behind Shapley values is that we wish to see the contribution of a specific feature when we include or exclude it from our data point for varying coalitions of features.

More formally, we can calculate the feature value importance as follows for a data point:

\begin{equation}\label{feature_value_importance_equation}
\phi_i^j = \sum_{S \subseteq N} \frac{\vert S \vert ! (n - \vert S \vert -1)!}{n!} (g(x'^j_i) - \hat{g}(x'^j_{-i})), 
\end{equation}

where $N$ is the set of features, $n$ \textbf{is the number of features}, $S$ is the subset, $g(x'^j_i)$ is the explanatory model evaluated when the feature is included and $g(x'^j_{-i})$ the explanatory model evaluated when the feature is not included. We refer to $\phi_i^j$ as a Shapley value.

We can specifically make a connection between the function we are trying to explain, $f(x)$, and the explanation function by noting that $\phi_0 = \EX[f(x)] = \frac{1}{d}\sum_{j=1}^d f(x_j)$, where $d$ is the number of data points. By setting all $\boldsymbol{x'^{j}_{i}}$ equal to 1 we obtain:

\begin{equation}\label{final_shap_equation}
f(x^j) = \hat{g}(x'^j) = \EX[f(x)] + \sum_{i=1}^{n}\phi_{i}^j,
\end{equation}

which implies that the value of a function at a given data point is equal to the global average of the function (i.e. $\EX[f(x)]$) plus the feature value importances we calculate for that data point.

We now explain what this entails practically. Say that we have a data point of the form ($n$, $\zeta$, $T$, $m_{z}$, $\psi$) = ($10^3$, 500, 50, 1, 500) and we are interested in determining the contribution of the temperature being 50 K in producing an abundance of, say, $10^{-6}$. What this entails is taking all subsets of the set of features. Two of these subsets might be: 

\begin{itemize}
    \item All of the original features
    \item All of the original features except the density
\end{itemize}

For the first of these subsets, we consider the change in the value of the explanatory model, $\hat{g}$, when we include and exclude the temperature value of $x_3 = 50$. ``Excluding'' simply means that we replace the 50 K with a randomly drawn temperature value from our dataset of temperatures. We then compute the feature explanation model when this temperature value is included and take the difference as seen in Equation \ref{feature_value_importance_equation}. For the second sample subset, we repeat this process except we always take a random value for the density as this is excluded from this subset. This is done for all subsets to calculate the feature importance for temperature.

However, observe that the calculation across all the subsets becomes computationally unfeasible as the number of features grows, with the number of coalitions growing exponentially. We employ SHAP \citep{SHAP_paper} to allow us to address this issue.  SHAP is particularly useful, as it approximates the Shapley values, greatly reducing the time taken to compute them. SHAP has been found to be the theoretically optimal means of calculating feature attribution \citep{SHAP_paper, TreeSHAP}. This is done through the use of the TreeSHAP algorithm \citep{TreeSHAP}. TreeSHAP is an algorithm that exactly computes the SHAP values for tree-based algorithms, such as XGBoost or random forests. One drawback of TreeSHAP is that it can \textbf{give} unintuitive explanations when the features are related \citep{molnar2022}. This is unlikely to be the case in this work, as we work with five physically unrelated physical features that we sample independently when we generate our data set.

We can also provide a ranking of the various features in terms of global feature importance. As Shapley values can be negative, this can be achieved by averaging the absolute value of all Shapley values for each feature across all datapoints. Formally, this is defined as: 

\begin{equation}\label{average_importance}
I_{i} = \frac{1}{d}\sum_{j=1}^{d} \lvert \phi_{i}^{j} \rvert,
\end{equation}

where $d$ is the number of data points and $I_{j}$ is the average absolute value of the $j$th feature. 

In principle, if we wished to compute the relative importances of the features we can do this by taking the above-mentioned average of the absolute values for a single feature and normalising this by the sum of the average of absolute values for all the features. We can then define the ``relative importance" for a feature $i$, $\hat{I_{i}}$, as:

\begin{equation}\label{relative_importance}
\hat{I_{i}} = \frac{\sum_{j=1}^{d} \lvert \phi_{i}^{j} \rvert}{\sum_{m=1}^{n}\sum_{j=1}^{d} \lvert \phi_{m}^{j} \rvert},
\end{equation}

where $n$ is the number of features. This quantity effectively gives us a fractional contribution of each feature to the average behaviour of the model. We summarise the relative importance of each parameter in predicting the outputs we consider in this work in Table \ref{relative_importance_table}.

\subsection{Implementation}

While the use of SHAP greatly reduces the time taken to obtain the Shapley values relative to calculating the Shapley values in full, this process is still likely to take long due to the time taken per evaluation of the forward model, i.e. UCLCHEM. Each evaluation of the forward model takes on the order of 1-2 minutes. This makes considering an ensemble of models with 100000 runs or more unfeasible. To circumvent this, we elect to train a statistical emulator to reproduce the results of UCLCHEM. If the emulator has a sufficiently high accuracy, then it is safe to assume it is able to capture the internal workings of the original code, which we wish to probe. We now discuss the emulator and how we build it. 

To train the emulator, we generated 120,000 points in parameter space using a Latin Hypercube sampling scheme \citep{LHS}, which was implemented using the Python surrogate modelling toolbox \citep{LHS_python}. Data points in parameter space were generated such that all values were in the ranges given in Table \ref{parameter_range_table}. For those features that spanned several orders of magnitude, we elected to sample in log-space. 

Each species' final log-abundance was used as the output of the algorithm. This was to ensure that all orders of magnitude were treated equally. The input parameters were also scaled to be in the range 0 to 1. All abundances less than $10^{-12}$ were set equal to $10^{-12}$ to ensure that the emulator was not being trained to learn what was effectively numerical noise. This limit was chosen because this is typically the lowest observed gas-phase abundance in the literature. We summarise the range of outputs for each species and ratio we consider in this work in Table \ref{output_abundances}.

An XGBoot regressor was trained for the emulation process \citep{XGBoost}. XGBoost is a gradient-boosted decision tree regressor. We used the Python implementation for XGBoost to train our model\footnote{https://xgboost.readthedocs.io/en/stable/index.html}. It was found that better performance was obtained if a separate emulator was trained for each species, as opposed to having one network trained to predict the final abundances of all 239 species in the network. While we trained an emulator for every species in the network, we only present the results of a handful of molecules in this work. We elected to train an XGBoost model instead of using a neural network as in \cite{Damien}, as XGBoost has been found to perform better on tabular datasets such as the one we consider, while also requiring less tuning \citep{xgboost_benchmarking}.

In order to find the best set of hyperparameters for each emulator, we utilised Bayesian Hyperparameter optimisation. Under this procedure, we tune the hyperparameters on a validation set and find the best combination of parameters that minimise the L2 loss. Unlike a grid-search approach to hyperparameter tuning, Bayesian optimisation uses the model performance on previous hyperparameter combinations to choose a next best option, thereby saving a considerable amount of time compared to a grid-search approach. XGBoost has five tunable hyperparameters that we varied using the Bayesian Optimisation Python library \citep{Bayesian_hyperparameter_optimisatoin}. We list the ranges over which we varied these in Table \ref{hyperparameter_table}. For integer hyperparameters, we would round to the nearest integer.

When evaluating the accuracy of each trained emulator, we considered both the L2 loss obtained obtained from the performance of the emulator on the test dataset as well as the $R^2$ coefficient. All emulators in this work had $R^2$ scores greater than 0.98.

\begin{table}
\begin{tabular}{lll}
\hline
\textbf{Hyperparameter} & \textbf{Range of values} & \textbf{Data type} \\ \hline
Maximum Depth           & (3, 100)                  & Integer            \\
Maximum features        & (0.8, 1.0)               & Float              \\
Learning Rate           & (0.01, 1.0)              & Float              \\
Number of Estimators    & (80, 150)                & Integer            \\
Sub-sample              & (0.8, 1)                 & Float              \\ \hline
\end{tabular}
\caption{Table of the hyperparameter ranges used when tuning the XGBoost regressor.}
\label{hyperparameter_table}
\end{table}

\section{Results}\label{sec:results}
We now look to consider a number of molecules of interest and explore how machine learning interpretability adds to our understanding of their equilibrium abundances in Phase 2. Note that all the molecules we will be considering will be gas-phase molecules, as they evaporate during the warm-up phase. We only considered a small number of molecules as a proof-concept for this method and provide the figures for these here. Figures for other molecules can be found in a dedicated repository.\footnote{https://github.com/Bamash/MLinAstrochemistry}.

\subsection{Molecules}
We begin by first considering individual molecules of interest to demonstrate what can be done with machine learning interpretability. We elect to consider three molecules: \ce{H2O}, \ce{CO} and \ce{NH3}. CO is considered as it is the most abundant molecule besides \ce{H2} and also plays a role in molecular gas cooling \citep{CO_cooling, CO_in_metal_poor_galaxies}. \ce{H2O} is of interest due to its high abundance in planetary system and of course its importance in the area of astrobiology \citep{water_Gensheimer}.  \ce{NH3} is speculated to be one of the main carriers of nitrogen and it often used as a tracer molecule of cold, dense clouds \citep{cold_cloud_survey, ammonia_caselli}.

\subsubsection{\ce{H2O}}
\begin{figure*}
\includegraphics[width=2.1\columnwidth]{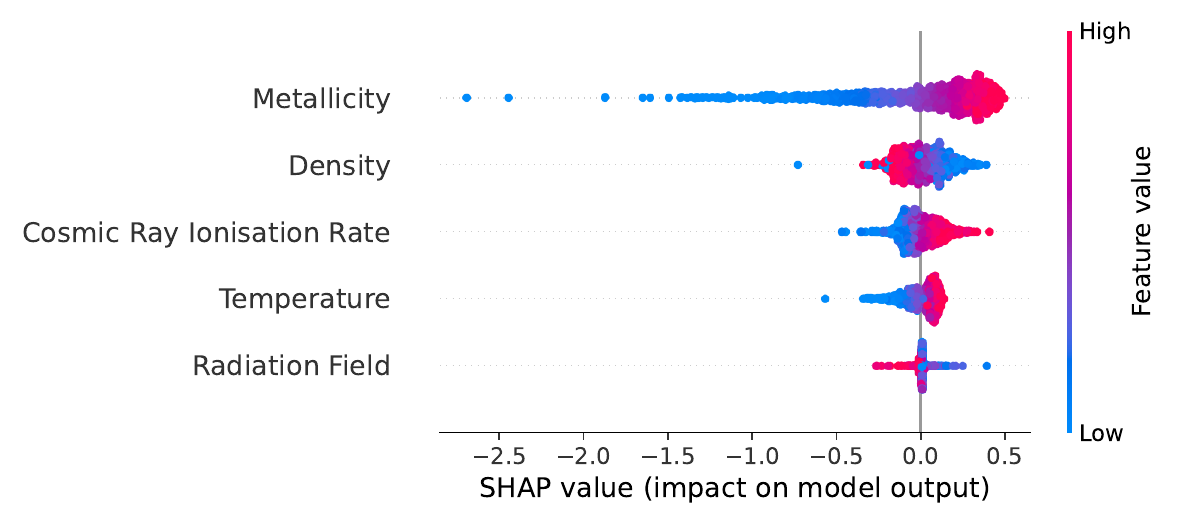}
\caption{A beeswarm plot of the various physical parameters demonstrating their relative importance in predicting the log-abundance of \ce{H2O}. The features are arranged from top to bottom in decreasing order of importance to the model output, which is measured by the mean of the absolute value of the SHAP value averaged across all predictions. Individual predictions are plotted along the horizontal axis according to their SHAP value, which indicates the difference in the value of the model output for that prediction relative to the global average. Furthermore, the points are colour-coded in terms of the size of the feature value relative to the range of values that the respective feature takes. We observe that metallicity ($\hat{I}_{m_z} = 0.54$) has the greatest impact followed by density ($\hat{I}_{n} = 0.17$), cosmic ray ionisation rate ($\hat{I}_{\zeta} = 0.15$), temperature ($\hat{I}_{T} = 0.17$) and radiation field ($\hat{I}_{\psi} = 0.02$).}
\label{H2O_dot_plot}
\end{figure*}

We now investigate the importance of the various physical parameters on the value of the abundance of \ce{H2O}. Figure \ref{H2O_dot_plot} is a beeswarm plot which is meant to serve as an information-dense qualitative summary of feature importances. \textbf{Each point in the beeswarm plot represents a data point from our test set.} The features are arranged from top to bottom in decreasing order of importance to the model output, which is measured by Equation \ref{average_importance}. Recall that the SHAP value measures the impact of each feature on the value of the prediction, relative to some baseline value, which is simply the global average, i.e. the average logarithm of the abundance. Along the horizontal axis, individual predictions are plotted in terms of their SHAP value. The points are colour-coded according to their value with the colour bar indicating the value relative to the range of values of that feature. A single colour bar is used for all the features in order to qualitiatively show the relationship between the feature value and the SHAP value. It is for this reason that the colour bar range is from ``Low" (indicating the lowest value of the respective feature) to ``High" (indicating the highest value that the feature can take. Furthermore, the vertical clustering of the points indicates the density of the points in a manner akin to a violin plot. We emphasise again that this plot is meant to help provide easy-to-use qualitative explanations for observers.

\begin{figure*}
\includegraphics[width=2.1\columnwidth]{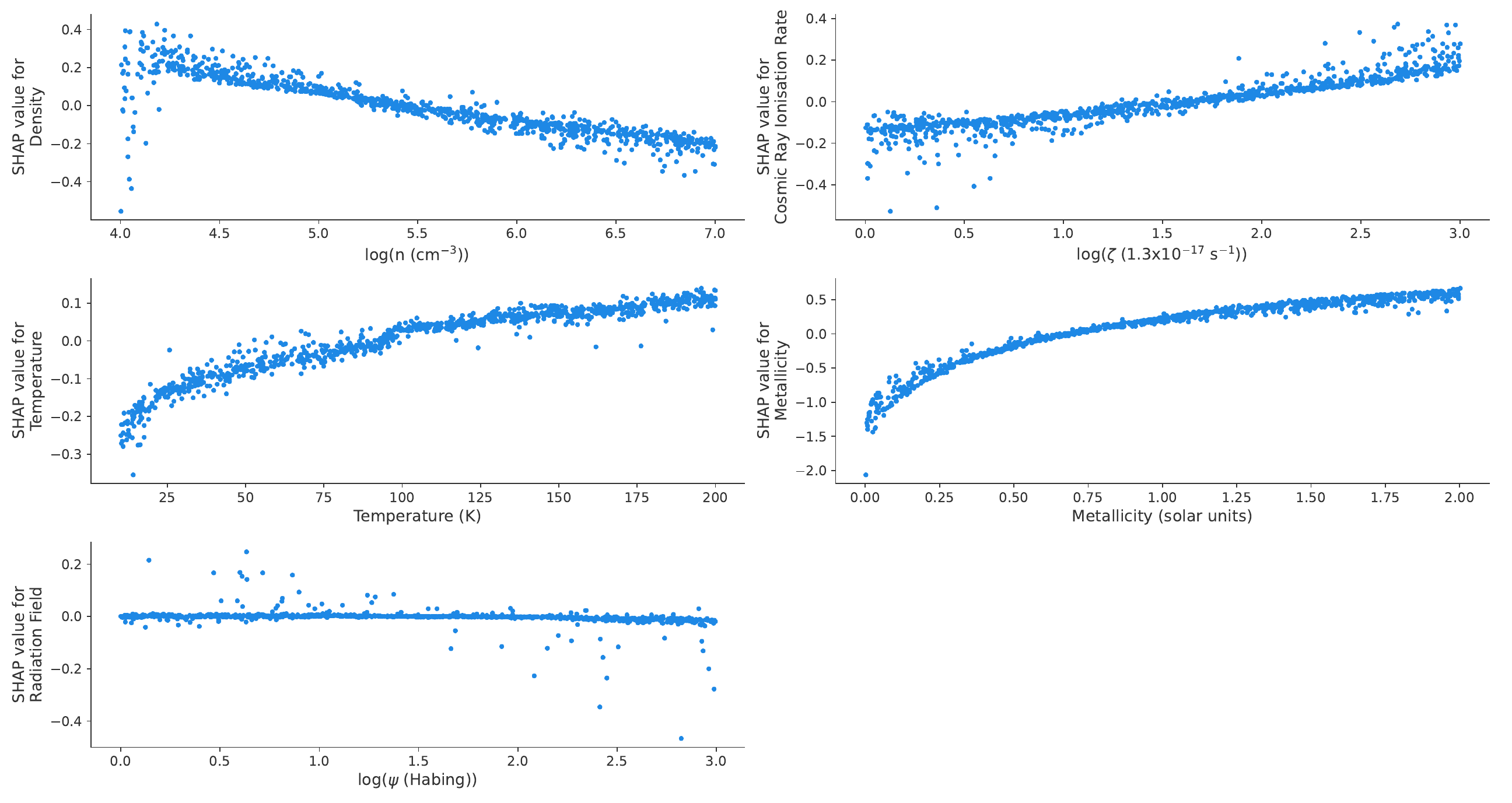}
\caption{A plot of the SHAP values as a function of the feature values used to predict the log-abundance of \ce{H2O}. Unlike the beeswarm plot, these SHAP dependence plots allow us to see the exact nature of the relationship between the feature value and SHAP value. Recall that the SHAP value tells us the difference in value between the average output value (log-abundance of the water). We see that the logarithms of density and the cosmic ray ionisation rate are roughly linear with respect to the SHAP value with the same being true for the temperature. For metallicity, we observe a significant decrease in the SHAP value for low metallicities, but this seems to level off for values greater than 1.}
\label{H2O_dependence_plot}
\end{figure*}

We also consider a more quantitative plot to delve deeper into some of the finer points of the beeswarm plot. This is useful if one wishes to consider the nature of the relationship between each feature and the log-ratio. While one can deduce that for temperature and metallicity the relationship is monotonic and increasing, this might still not be enough. It is for this reason that we can plot dependence plots such as Figure \ref{H2O_dependence_plot} which plots the SHAP value for all the variables as a function of the individual variable. Notice that in the plot for each feature $i$, the SHAP value corresponds to the importance of only that feature, $\phi_i^j$, for a point $j$. Effectively, these dependence plots gives us the marginal contribution of each feature $i$ to the output.

\begin{figure*}
\includegraphics[width=2.1\columnwidth]{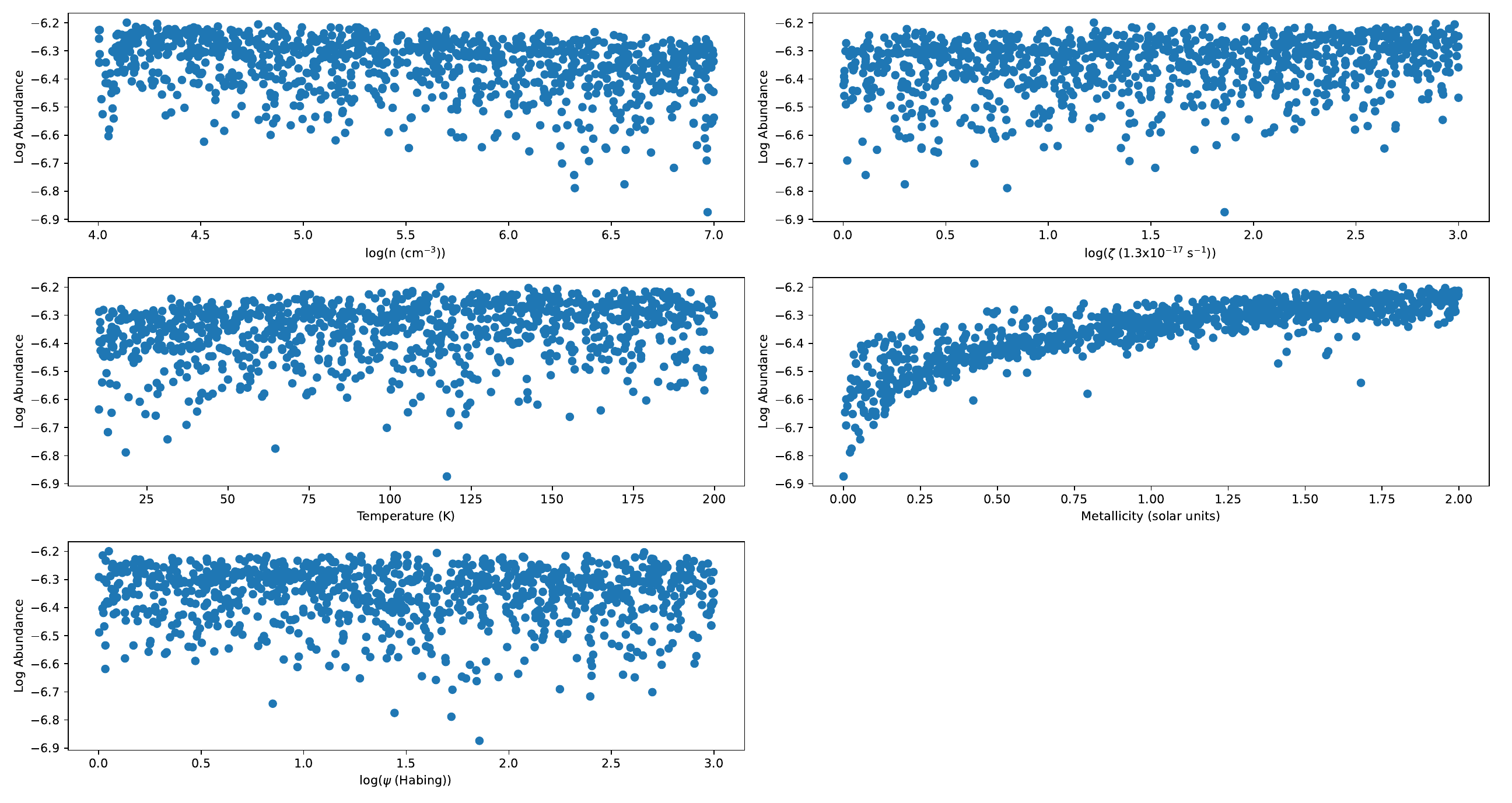}
\caption{A plot of the log-abundance of \ce{H2O} as a function of the various features. To calculate the log-abundance for a given data point, we needed to sum up the importance values of each feature for that data point. We observe that only metallicity maintains a clear trend. For the other features, we have no discernible trend which can be attributed to the feature importances nullifying each other.}
\label{H2O_native_plot}
\end{figure*}

We can also consider the relationship between the abundance of water (instead of the SHAP value) as a function of each of the features. This is plotted in Figure \ref{H2O_native_plot}. Notice that in order to compute the abundance, we must utilise Equation \ref{final_shap_equation}. This means that to compute the abundance we must add the mean log-abundance of water, $\phi_0 = \EX[f(x)]$, to the SHAP values of each of the features for that data point. As a result of the explanatory model being linear in nature, we do not see the same relationships in Figure \ref{H2O_native_plot} and in fact observe that there is no relationship between all the parameters besides metallicity and the log-abundance. This is because many of the SHAP importances cancel each other out. Only metallicity still has a noticeable relationship with the log-abundance when we add up the importances of all the parameters.

However, in the interest of better understanding the impact of each parameter's individual relationship with the log-abundance, we consider the marginal effects in Figure \ref{H2O_dependence_plot}. We would like to emphasise that it is still useful to consider the marginal effects. While we consider a wide range of physical conditions, many observational and modelling exercises relating to tracers will be far more restrictive in their parameter ranges as well as the number of varying parameters. A typical observational environment will not contain the parameter ranges we consider here. It is precisely for these tasks that this methodology will be useful. We observe that the relationships between thes variables and the log-abundance of \ce{H2O} are mostly monotonic. However, we will only consider the impact of metallicity, as this has the strongest impact on the model output. The SHAP values for the other features range between -0.4 and 0.4 in log-abundance space which corresponds to factors of 2.5 relative to the average water abundance. Throughout this work, we will only consider features whose SHAP values exceed 1 in log-abundance space. It is clear that metallicity will play a significant role in the abundance of water. While there exists some debate as to what fraction of the ISM oxygen abundance is present in water \citep{water_in_star_forming_regions}, a decrease in the metallicity will result in a decrease in the amount of oxygen, which in turn will mean that less water will be formed, due to greater competition for the little oxygen present. On the other hand, a large amount of oxygen will result in the opposite effect, to an extent. Water has several destruction pathways that impose an upper limit on how much of it is formed in the gas-phase, regardless of how much oxygen is present.

\subsubsection{CO}
\begin{figure*}
\includegraphics[width=2.1\columnwidth]{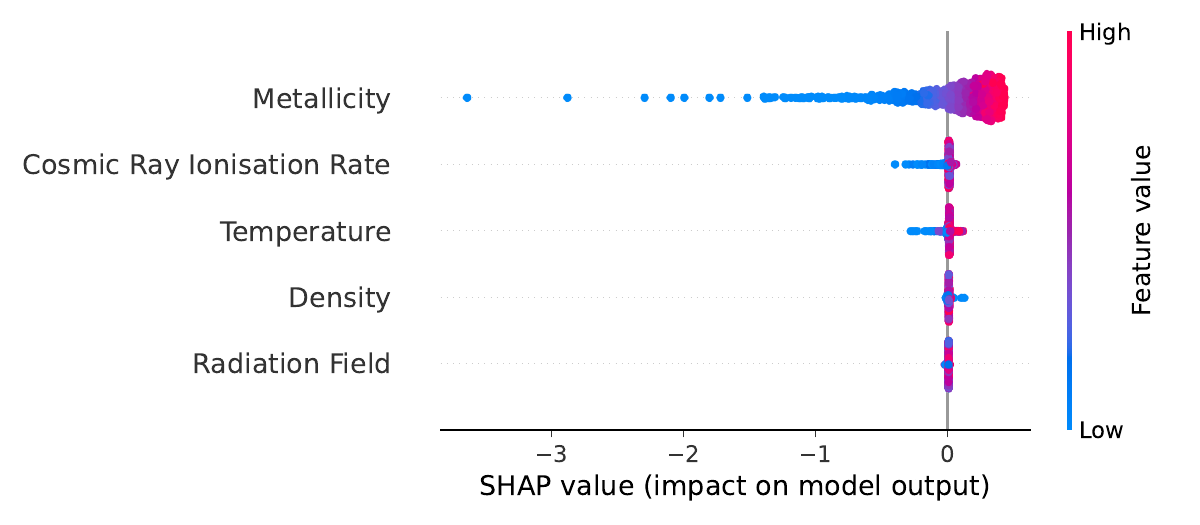}
\caption{A beeswarm plot of the various physical parameters demonstrating their relative importance in predicting the log-abundance of \ce{CO}. We observe that metallicity is the only parameter with a significant influence on the value ($\hat{I}_{m_z} = 0.91$) with the other parameters not being very useful predictors ($\hat{I}_{n} = 0.01$, $\hat{I}_{\zeta} = 0.04$, $\hat{I}_{T} = 0.04$ and $\hat{I}_{\psi} = 0.00$).}
\label{CO_dot_plot}
\end{figure*}

\begin{figure*}
\includegraphics[width=2.1\columnwidth]{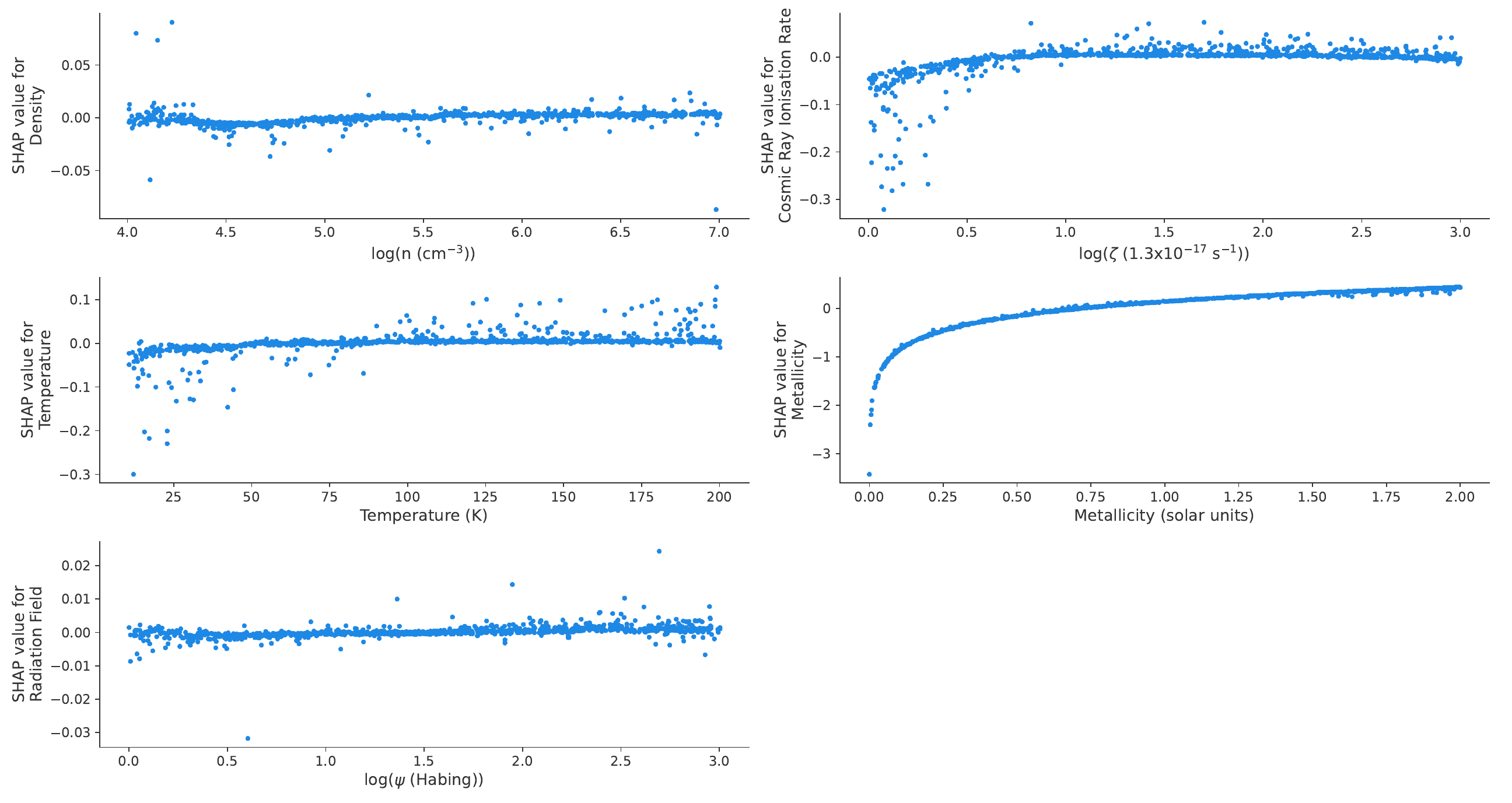}
\caption{A plot of the SHAP values for the various features (besides the radiation field) as a function of the feature values used to predict the log-abundance of \ce{CO}. As was observed in the beeswarm plot, only metallicity has a significant effect on the abundance. For low metallicities, we observe a large decrease in the SHAP value. The SHAP value monotonically increases with metallicity, eventually levelling off for values greater than 1.}
\label{CO_dependence_plot}
\end{figure*}

\begin{figure*}
\includegraphics[width=2.1\columnwidth]{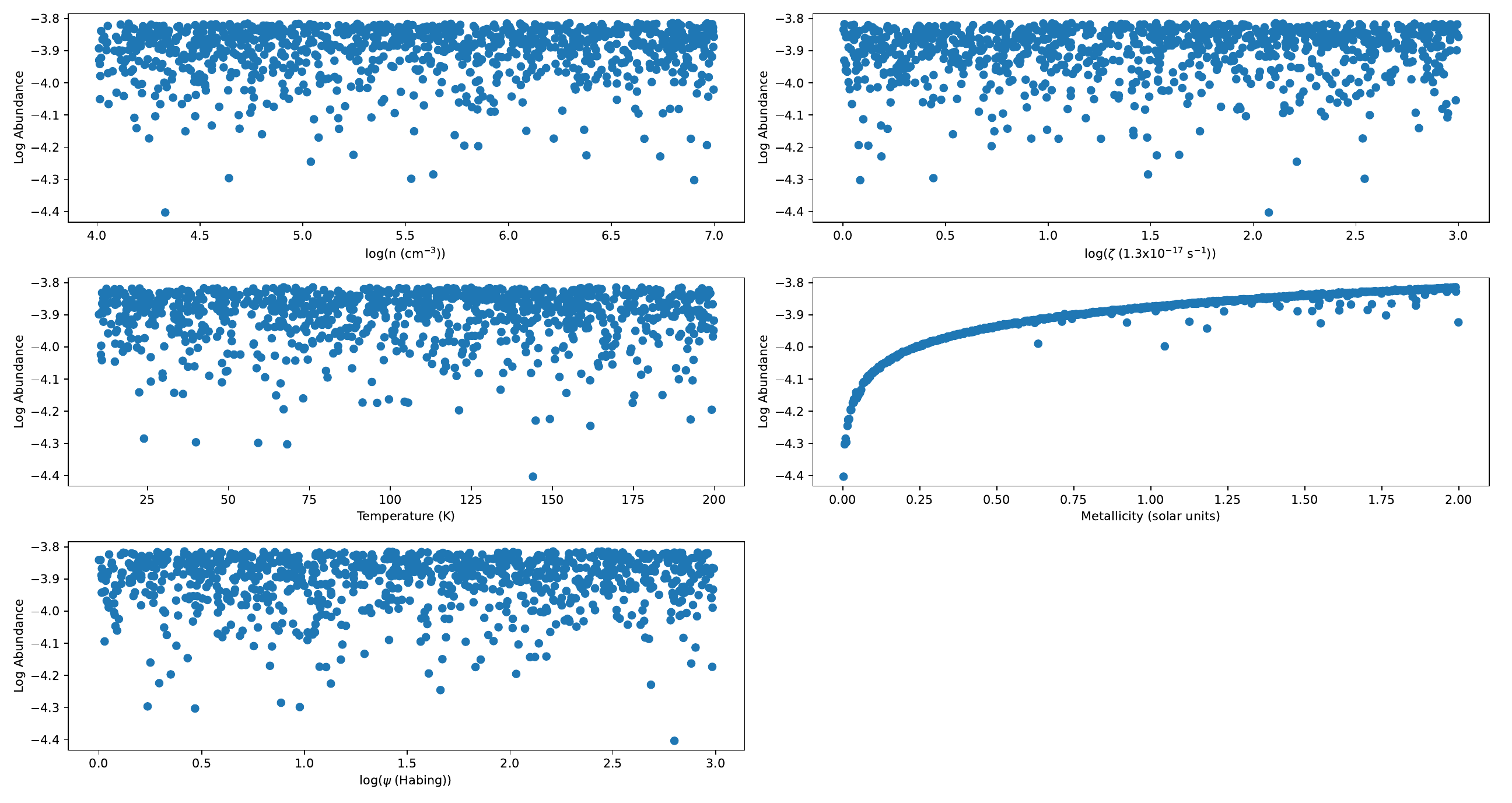}
\caption{A plot of the log-abundance of \ce{CO} as a function of the various features. To calculate the log-abundance for a given data point, we needed to sum up the importance values of each feature for that data point. Only metallicity maintains a clear trend compared to Figure \ref{CO_dependence_plot}. For the other features, we have no discernible trend. This is due to the marginal feature importances nullifying each other.}
\label{CO_native_plot}
\end{figure*}

Carbon monoxide is an important molecule to consider in astrochemistry. Not only is it an important molecule in the context of grain-surface chemistry and the formation of various complex organic molecules, but it also plays a significant role in gas-phase chemistry. In particular, it is often considered a molecular gas coolant at low temperatures and densities \citep{CO_cooling, CO_in_metal_poor_galaxies}. We are interested in considering how the various parameters we are changing influence its abundance. Figure \ref{CO_dot_plot} is a beeswarm plot of the various features and shows that only the metallicity plays a strong role in determining the final CO abundance, which has an $\hat{I_i}$ of 0.91. In order to investigate the exact nature of the relationship, we plot the SHAP dependence plots in Figure \ref{CO_dependence_plot}. We observe an interesting relationship between the metallicity and the CO abundance that is monotonic in nature. We do not observe any notable relationships between its log-abundance and the other parameters, so we only focus on metallicity for now. We observe that for very low metallicities the CO abundance ends up being almost 2 orders of magnitude lower than the `average' value due to the marginal effect of the metallicity. In Figure \ref{CO_native_plot} we plot the log-abundance of CO as a function of each of the parameters. As we discussed for \ce{H2O} before, to compute the CO abundance we must add the contributions of all the features. As a result of this, only the metallicity appears to have a strong effect on the log-abundance.

Work has been done to consider the impact metallicity on CO. In \cite{CO_in_metal_poor_galaxies}, this was considered in the context of metal-poor galaxies. Here they wished to consider to what extent CO, known to be a coolant in metal-rich galaxies, could also serve the same role in metal-poor ones. It was found that there was significant CO depletion in metal-poor galaxies due to photodissociation. This is unlikely to be the case here as the radiation field is found to be the least influential parameter. The radiation field is only likely to be effective in photodissociation when the density is very low and the radiation field itself is high, which will only be the case for a small number of parameter combinations. We must consider other reasons for the importance of metallicity.

Within the UCLCHEM code, the metallicity parameter is a scale factor that scales all elemental abundances of elements heavier than helium by the same factor. This means that as the metallicity parameter is reduced, the abundances of some elements will be reduced so there will be greater competition for them. This results in the abundances of species dropping, as there is simply less of their constituent elements. In the case of CO, we know from Table \ref{elemental_abundances}, that there is less C than O which means reducing the metallicity results in C becoming more scarce. It is for this reason that at metallicities close to zero the final abundance of CO drops by 2 orders of magnitude.

\subsubsection{\ce{NH3}}
\begin{figure*}
\includegraphics[width=2.1\columnwidth]{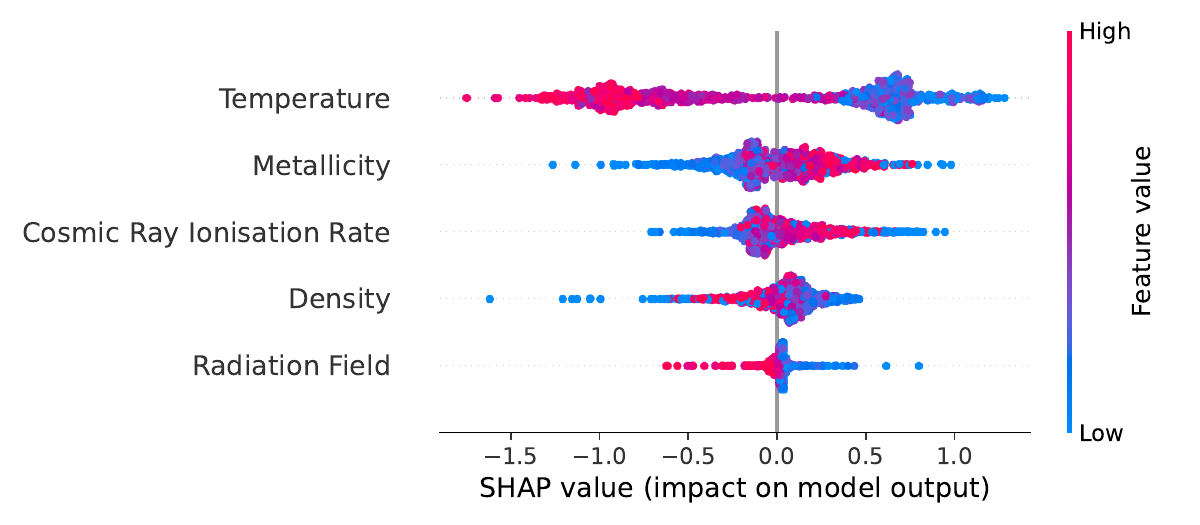}
\caption{A beeswarm plot of the various physical parameters demonstrating their relative importance in predicting the log-abundance of \ce{NH3}. We observe that temperature has the largest impact with ($\hat{I}_{T} = 0.54$). The temperature relationships does not seem to be monotonic. The next most important features are metallicity ($\hat{I}_{m_z} = 0.17$), followed by the cosmic ray ionisation rate ($\hat{I}_{\zeta} = 0.14$), density ($\hat{I}_{n} = 0.11$), and the radiation field ($\hat{I}_{\psi} = 0.03$), with the first three also not having monotonic relationships with the SHAP value.}
\label{NH3_dot_plot}
\end{figure*}

\begin{figure*}
\includegraphics[width=2.1\columnwidth]{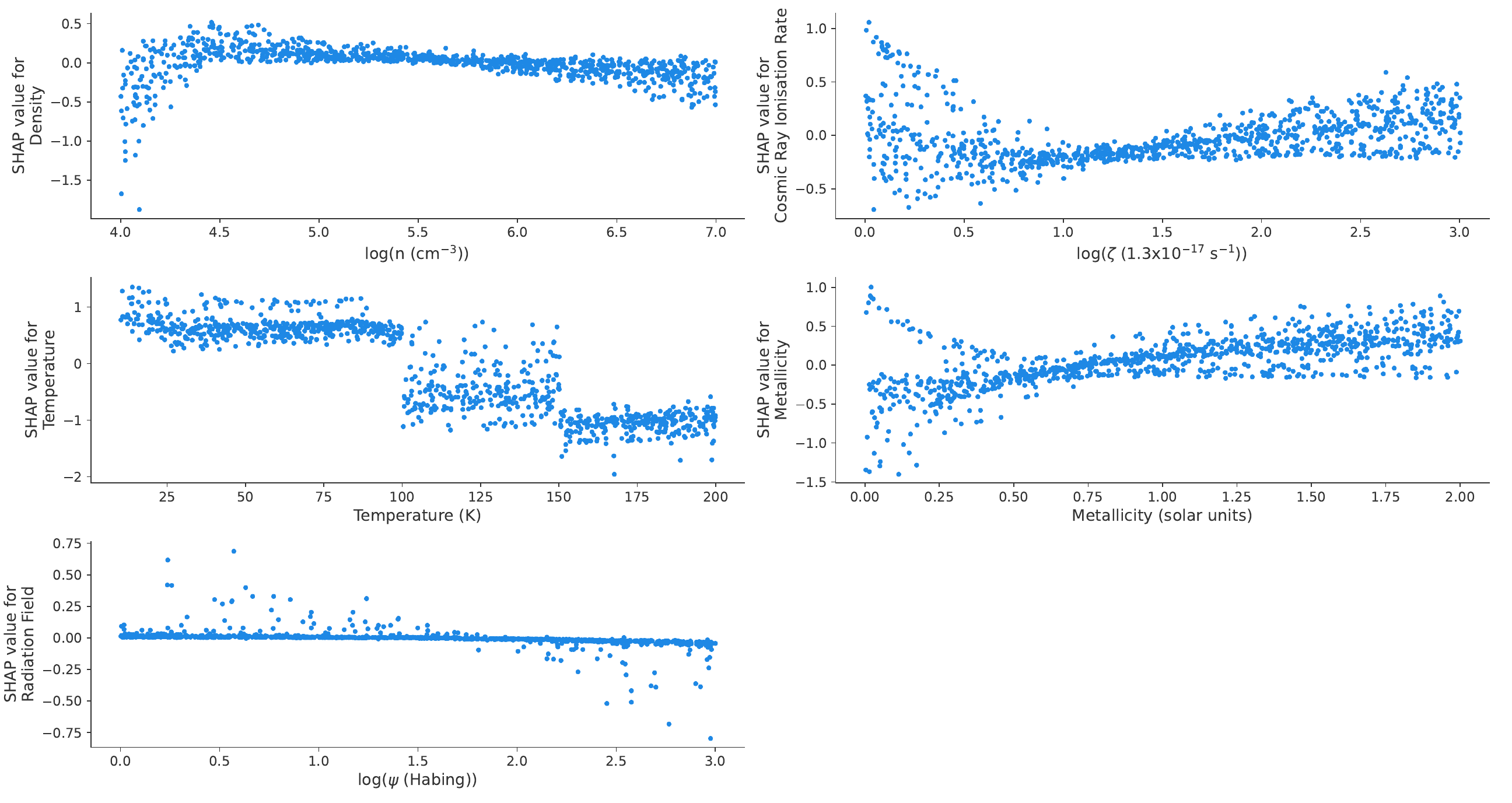}
\caption{A plot of the SHAP values for the various features (besides the radiation field) as a function of the feature values used to predict the log-abundance of \ce{NH3}. We observe that temperature has an interesting relationship with the SHAP value. What we observe is that there exist three separate temperature regimes under which the final abundance is relatively constant. The abundance does show some non-monotonic variance with respect to the other features, but most of these are within 0.5 of the average value (or a multiplicative factor of 3).}
\label{NH3_dependence_plot}
\end{figure*}

We now consider ammonia, which is considered one of the significant sources of nitrogen in the interstellar medium. The beeswarm plot in Figure \ref{NH3_dot_plot} shows the ranking of the 5 features in terms of their relative importance. We consider the nature of the relationship through the use of the SHAP dependence plots in Figure \ref{NH3_dependence_plot} with only temperature being found to have a consistently significant relationship with the log-abundance. Parameters such density and cosmic ray ionisation rate may have individual points with large SHAP values but these are low in frequency compared to the tens of thousands of points plotted, which is why we do not discuss them further. The dependence on metallicity is not as strong as for \ce{H2O} and CO in terms of the tailing-off trend as the metallicity approaches zero. This is likely to be due to the highly non-linear nature of the chemistry. Despite the metallicity decreasing it is likely that there is some reaction that is compensating for the decrease in \ce{NH3} such that more of the now limited nitrogen is now hydrogenated.

\begin{figure*}
\includegraphics[width=2.1\columnwidth]{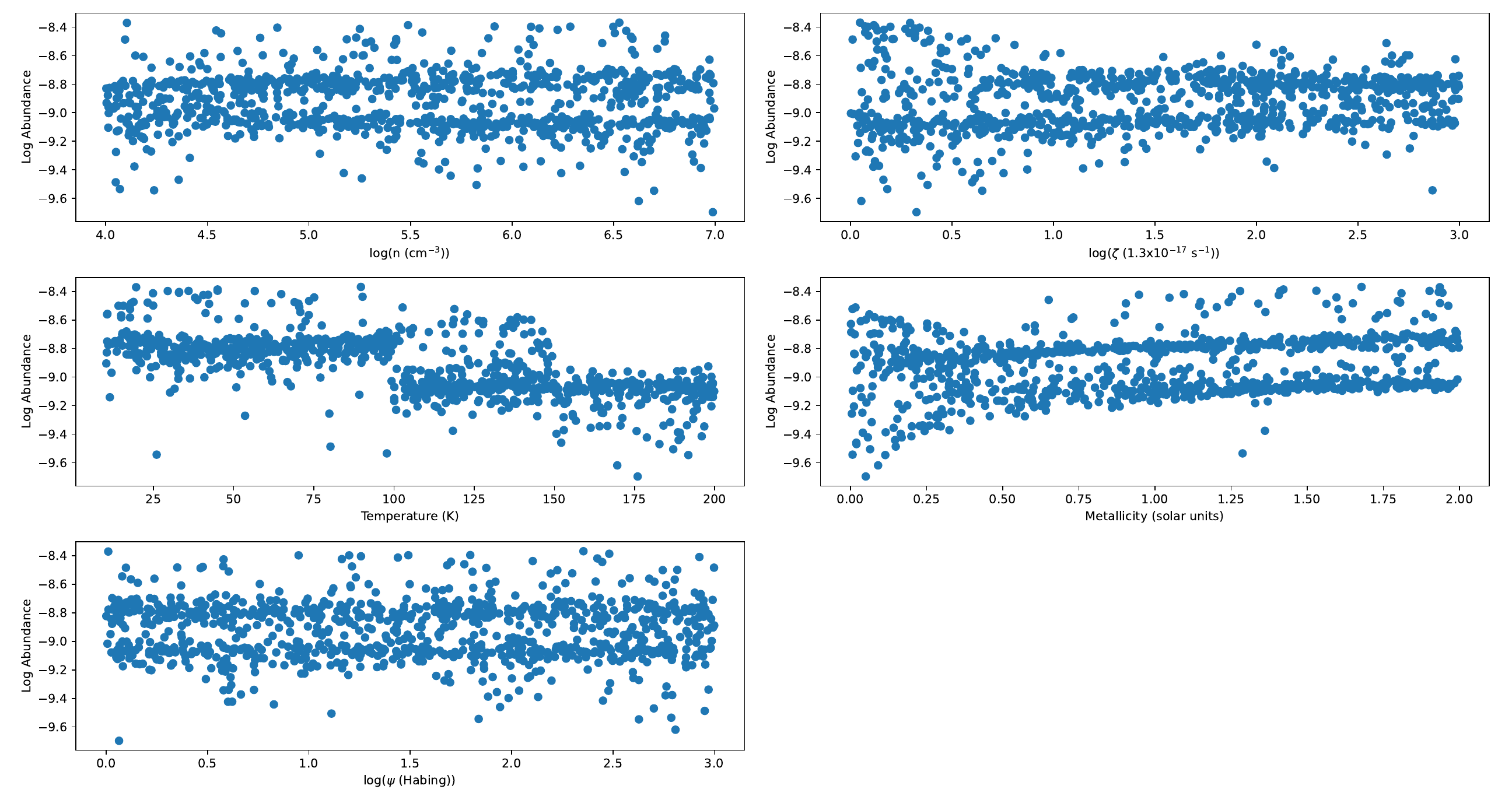}
\caption{A plot of the log-abundance of \ce{NH3} as a function of the various features. To calculate the log-abundance for a given data point, we needed to sum up the importance values of each feature for that data point. We observe that only temperature maintains a clear trend relative to what we observed in Figure \ref{NH3_dependence_plot}. However, we now appear to have something closer to a two-temperature regime rather than a three-temperature one. For the other features, we have no discernible trend which can be attributed to the feature importances nullifying each other.}
\label{NH3_native_plot}
\end{figure*}

The dependence on temperature is quite interesting, as we notice that there are two separate temperature ranges over which the abundance takes a different constant value, with the cutoff temperature being 100 K. This is also seen in though Figure \ref{NH3_native_plot} which is a plot of the abundances as a function of the individual parameters seems to indicate two regimes. We deduce that these different regimes are related to the chemistry surrounding NH$_3$ being very different at these stages. The other parameters are of less importance as the deviation from the average NH$_3$ log-abundance is within about 0.5, or a factor in 3 in actual abundance. The non-temperature parameters cancel each other out in terms of their contributions when these are added together.

\begin{figure*}
\includegraphics[width=2.1\columnwidth]{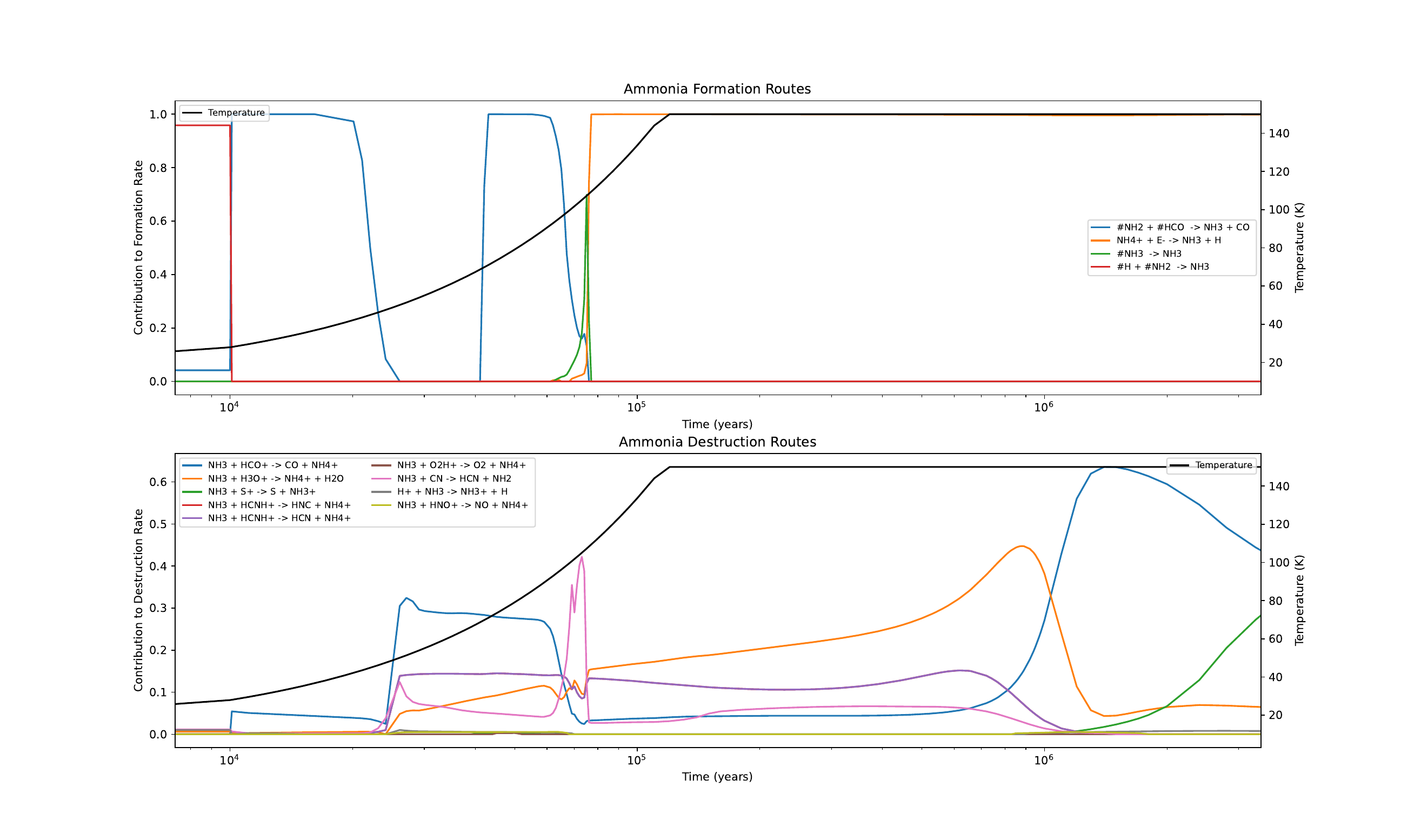}
\caption{Top: Plot of the fractional contribution of various ammonia formation routes that contribute to 99\% of the \ce{NH3} formation at each time. The temperature as a function of time is also plotted. Bottom: Plot of the fractional contribution of various ammonia destruction routes that contribute to 99\% of the \ce{NH3} at each moment in time. We only considered the top reactions that contributed to 99 \% of the creation or destruction to limit the number of lines we would have to plot. }
\label{NH3_reaction_rate_plot}
\end{figure*}

We can investigate the temperature dependence by considering the relative rates of formation and destruction of ammonia at specific points in time. Figure \ref{NH3_reaction_rate_plot} plots the fractional contributions of the various formation and destruction routes of \ce{NH3} for an instance where the peak temperature is 160 K. We only considered the top reactions that contributed to 99\% of the creation or destruction of \ce{NH3}. The main \ce{NH3} formation routes are: 

\begin{equation}
\ce{\#H + \#NH2 -> NH3} 
\end{equation}

\begin{equation}
\ce{NH4+ + e^{-} -> NH3 + H} \\ 
\end{equation}

\begin{equation}
\ce{\#NH3 -> NH3 (UV and CR desorption)} \\
\end{equation}

\begin{equation}
\ce{\#NH2 + \#HCO -> NH3 + CO} \\
\end{equation}

Some of the main destruction mechanisms throughout the hot core phase are: 

\begin{equation}
\ce{H3+ + NH3 -> NH4+ + H2} 
\end{equation}

\begin{equation}
\ce{NH3 + HCO+ -> CO + NH4+}
\end{equation}

\begin{equation}
\ce{NH3 + H3O+ -> NH4+ + H2O}
\end{equation}

\begin{equation}
\ce{NH3 + CN -> HCN + NH2} 
\end{equation}

\begin{equation}
\ce{NH3 + HNO+ -> NO + NH4+} 
\end{equation}

\begin{equation}
\ce{H+ + NH3 -> NH3+ + H} 
\end{equation}

\begin{equation}
\ce{NH3 + HCNH+ -> HCN + NH4+} 
\end{equation}

\begin{equation}
\ce{NH3 + HCNH+ -> HNC + NH4+} 
\end{equation}

\begin{equation}
\ce{NH3 + S+ -> NH3+ + S} 
\end{equation}

When the peak temperature is reached, the only formation reaction left is the gas-phase electron addition reaction. This is due to high temperature making grain-surface chemistry untenable, as most of the available grain-surface material has evaporated. In the gas-phase, the destruction routes are still active and recycle some of the gas-phase \ce{NH3} and turn it back into \ce{NH4+}, but some of it goes on to form HCN and other species, resulting in the eventual decrease in the NH$_3$ abundance. This is more severe for higher final hot-core temperatures as these destruction reactions see their rates increase, resulting in even lower final \ce{NH3} gas-phase abundances.

\subsection{Molecular ratios}
While species may serve as useful tracers for specific energetic processes under certain density and temperature conditions, it is often more useful to consider intensity ratios between different molecules, espeially in extragalactic environments \citep{Viti_2017, Imanishi_tracers, Josh_paper}.  Tracer ratios are often considered in observations to cancel out the beam filling factor. The two tracer ratios we consider are HCN/HNC and HCN/CS, both of which have been extensively studied in the literature. The former is considered a good tracer of temperature and the latter a dense gas tracer.

\subsubsection{HCN/HNC} 
We begin by considering the ratio of the abundances of HCN to HNC. The ratio of these two molecules has been extensively studied and has also been subject to a considerable amount of debate. These two molecules are of great interest, due to their high abundances, their excitation conditions the areas in which they form as well as the proximity of their transitions in frequency space \citep{Pety, HCN_HNC}. Recently, this intensity ratio was suggested as a potential chemical thermometer for the ISM \citep{HCN_HNC}. 

\begin{figure*}
\includegraphics[width=2.1\columnwidth]{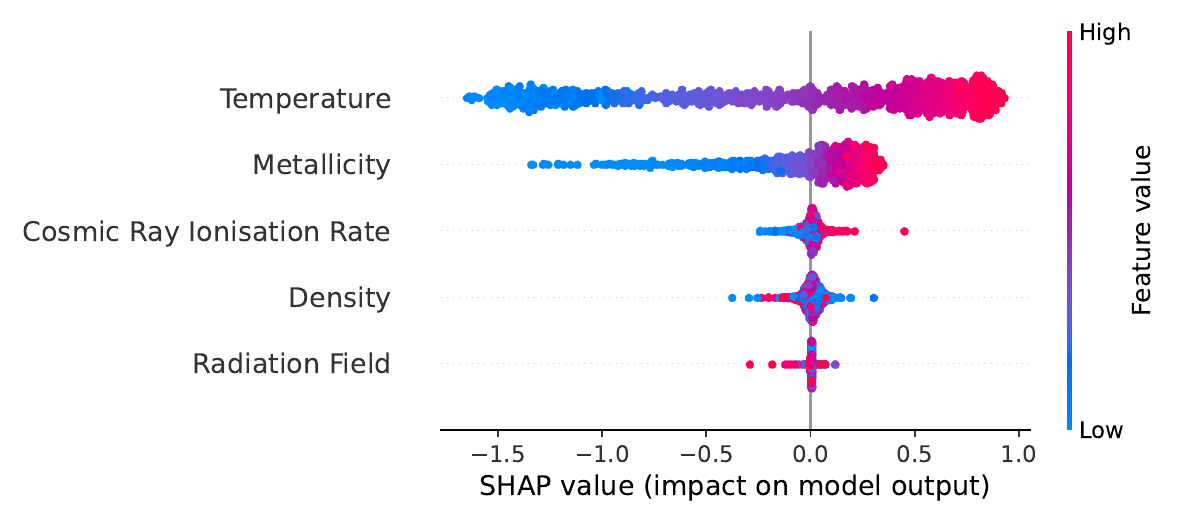}
\caption{A beeswarm plot of the various physical parameters demonstrating their relative importance in predicting the log-ratio of HCN to HNC. We observe that temperature has the largest impact on the model output with ($\hat{I}_{T} = 0.70$). The fact that temperature is the most important feature is hardly surprising given that this ratio is seen as a thermometer. The next most important features are metallicity ($\hat{I}_{m_z} = 0.24$), followed by the cosmic ray ionisation rate ($\hat{I}_{\zeta} = 0.03$), density ($\hat{I}_{n} = 0.03$) and the radiation field ($\hat{I}_{\psi} = 0.00$), with the first three also not having monotonic relationships with the SHAP value.}
\label{HCN_HNC_dot_plot}
\end{figure*}

\begin{figure*}
\includegraphics[width=2.1\columnwidth]{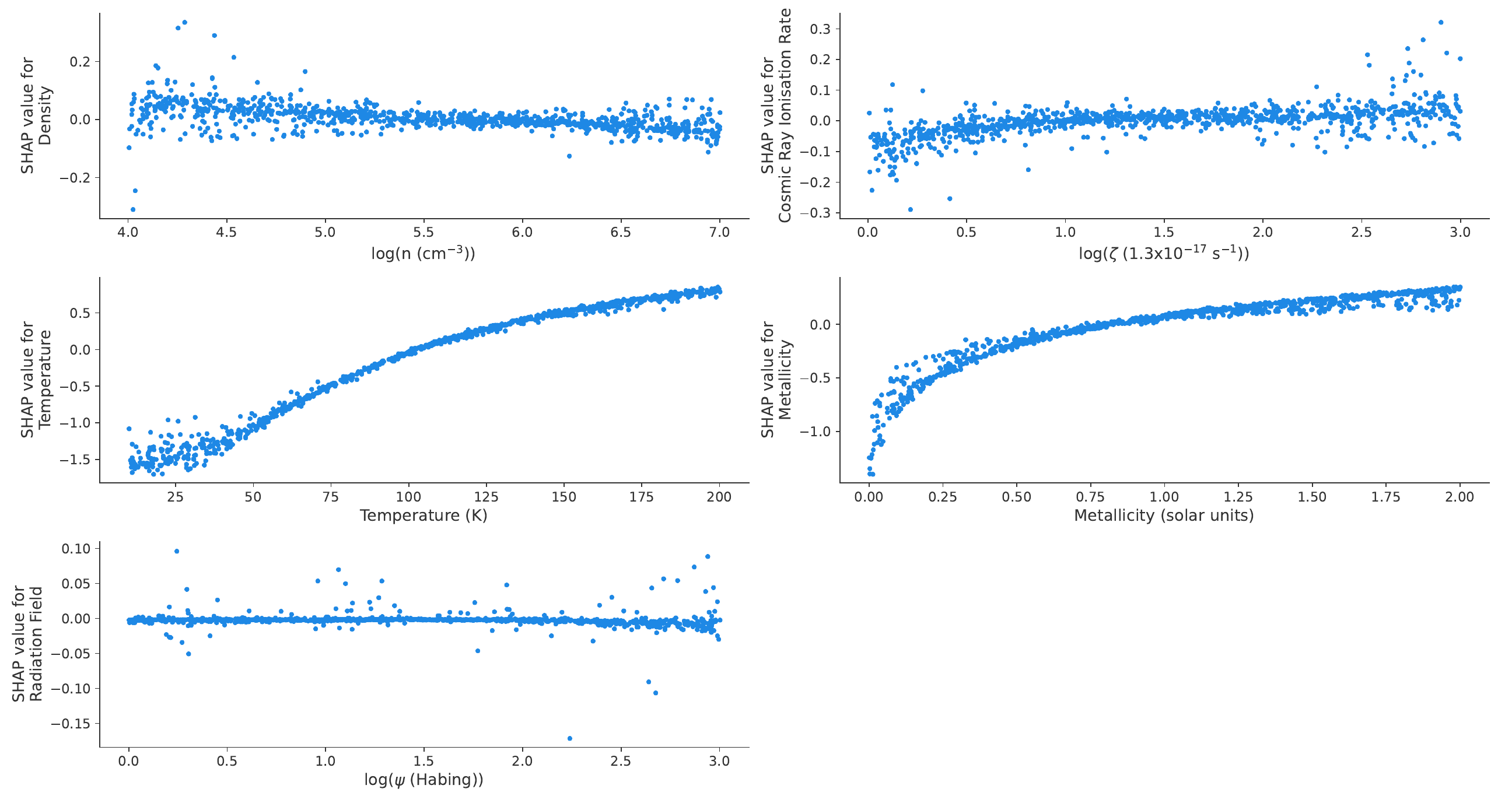}
\caption{A plot of the SHAP values for the various features (besides the radiation field) as a function of the feature values used to predict the log-ratio of HCN to HNC. We observe that temperature has an interesting relationship with the SHAP value with there being two regimes under which the ratio increases at different rates. This is in line with what was observed in \protect\cite{HCN_HNC} and was approximated there as a two-part linear function. The relationship between the SHAP value and metallicity is similar to what we observed in other molecules. }
\label{HCN_HNC_dependence_plots}
\end{figure*}

\begin{figure*}
\includegraphics[width=2.1\columnwidth]{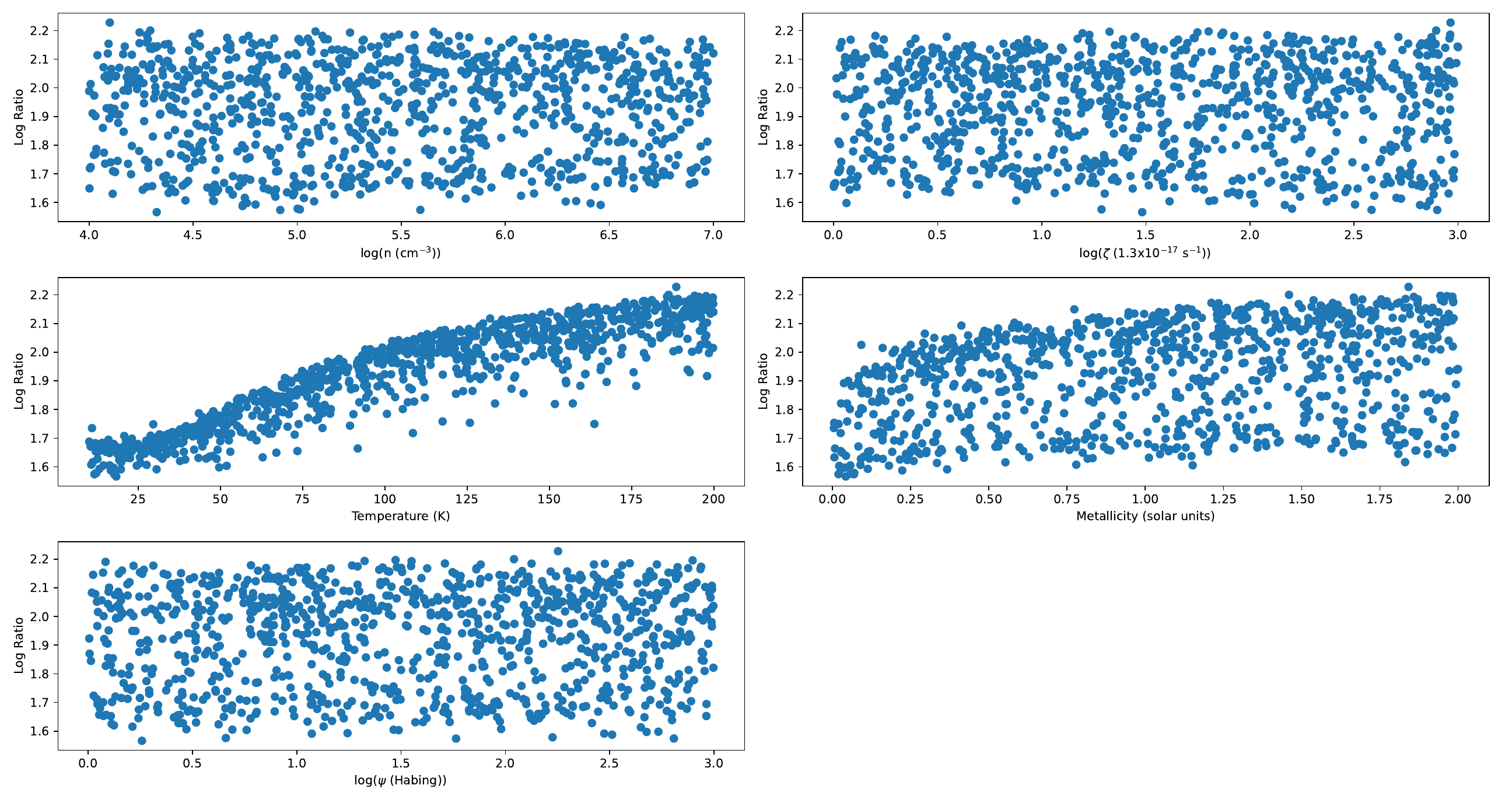}
\caption{A plot of the log-abundance of HCN/HNC ratio as a function of the various features. To calculate the log-ratio for a given data point, we needed to sum up the importance values of each feature for that data point. We observe that only temperature maintains a clear trend relative to what we observed in Figure \ref{HCN_HNC_dependence_plots}. For the other features, we have no discernible trend which can be attributed to the feature importances nullifying each other.}
\label{HCN_HNC_native_plots}
\end{figure*}

In Figure \ref{HCN_HNC_dot_plot}, we observe that temperature is indeed the most important feature. We observe that only the temperature and metallicity have significant impacts on the value of the ratio with relative importance values of $\hat{I_T} = 0.7$ and $\hat{I_{m_z}} = 0.24$. The other parameters do not have much influence on the log-abundance, so will not be discussed. Looking at the dependence plot for temperature further in Figure \ref{HCN_HNC_dependence_plots}, we observe that the log-ratio increases monotonically with temperature, with there appearing to be two different temperature regimes judging by the change in gradient throughout the curve, which is also evident in Figure \ref{HCN_HNC_native_plots} which is a plot of the log-ratio against the features. 

\begin{figure*}
\includegraphics[width=2.1\columnwidth]{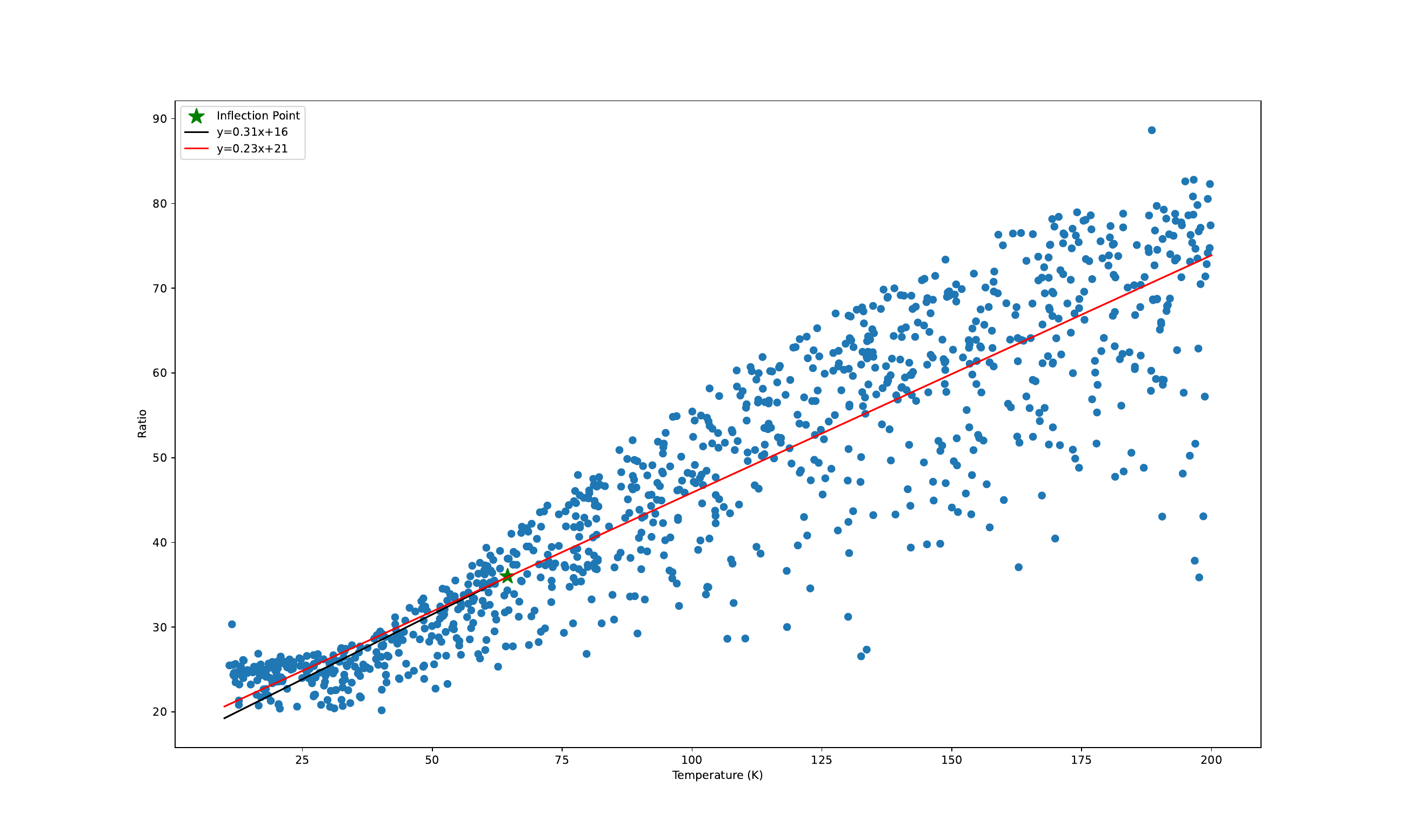}
\caption{Scatter plot of the ratio (note: not the log-ratio) as a function of the temperature. We continue to observe an inflection point at 65 K and fit a two-part linear function. Below 65 K, the trend line is $y = 0.31x + 16$ (black) and above it is $y = 0.23x + 21$ (red). For the sake of clarity, we have included the entirety of the second part of the red linear function to make the change in gradient easier to see.}
\label{temperature_plot}
\end{figure*}

Figure \ref{temperature_plot} is a plot of the ratio (as opposed to the log-ratio) against the temperature. We fit a two-part linear function to the data. The presence of two regimes is in agreement with the literature \citep{Graninger, HCN_HNC}. In \cite{HCN_HNC}, the relationship between the temperature and the ratio was described with a two-part linear function, which is what we roughly observe. The two isomers are formed in roughly equal proportions through the dissociative recombination of \ce{HCNH+} \citep{HCN_HNC_formation}. As such, any deviation in the ratio from a value of 1 can be attributed to the destruction routes. The main ones considered in the literature are: 

\begin{equation}
\ce{HNC + H -> HCN + H} 
\end{equation}

\begin{equation}
\ce{HNC + O -> NH + CO} 
\end{equation}

The pre-established energy barriers for both of these reactions have been questioned (see \cite{Graninger} for a full discussion of this). We update these values in line with \cite{HCN_HNC} and \cite{Graninger} to be 200K and 20 K respectively. The first reaction is particularly dominant at high temperature, where we have a large abundance of atomic H, whereas the second reaction is more dominant at low temperatures.

\begin{figure*}
\includegraphics[width=2.1\columnwidth]{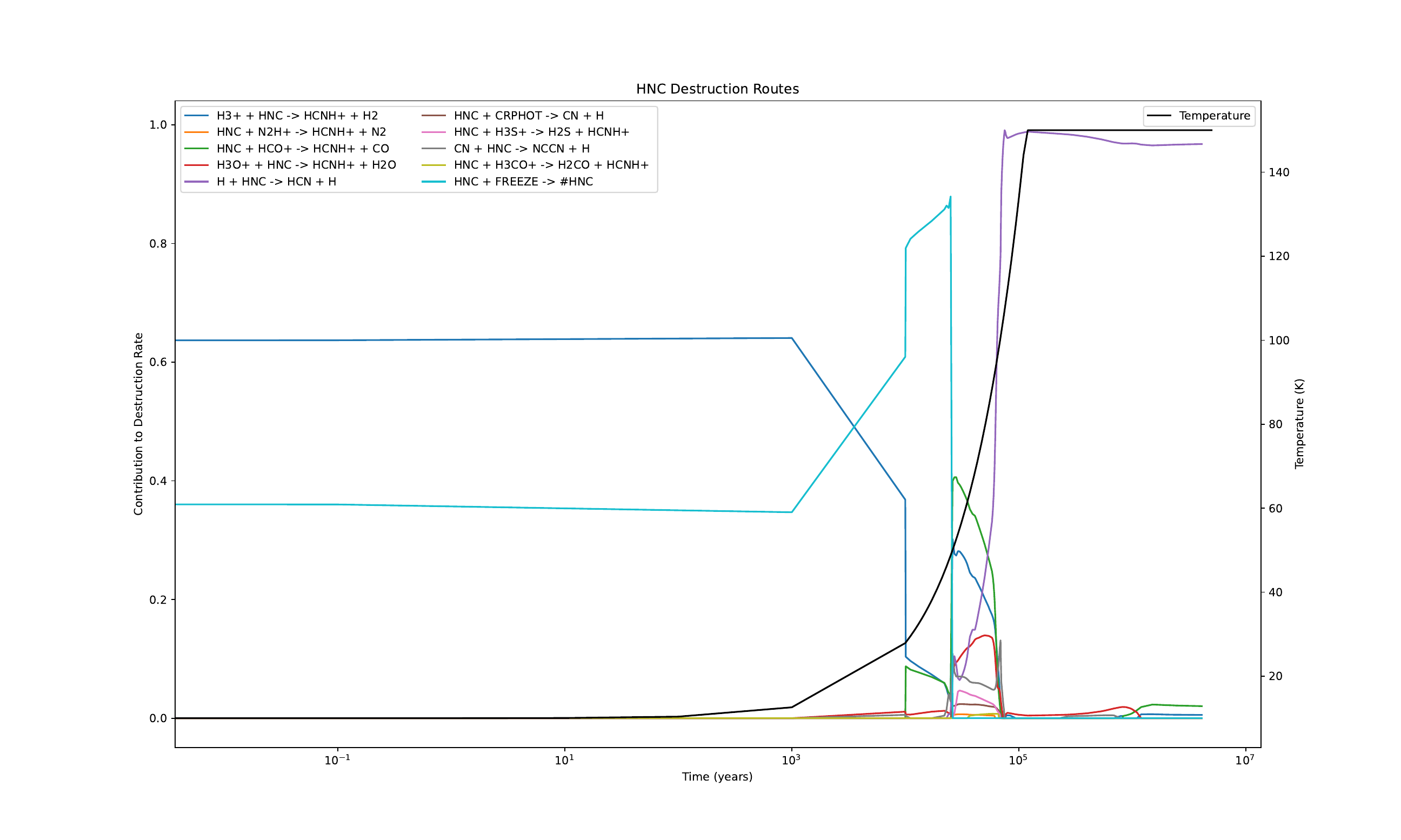}
\caption{Plot of the fractional contribution of various routes that contribute to 99\% of the \ce{HNC} destruction as a function of time. The temperature as a function of time is also plotted. We observe that for low temperatures, the main sources of gas-phase HNC destruction are \ce{H3+ + HNC -> HCNH+ + H2} as well as freeze-out onto the grains, which runs contrary to our expectations of the reaction \ce{HNC + O -> NH + CO} playing a dominant role. As the temperature increases we observe that the main destruction mechanism is the isomerisation reaction \ce{H + HNC -> HCN + H}. Note that the increase in the fractional contribution of the freeze-out reaction after $10^3$ years is not due to the increase in temeprature, but rather simply numerical as the other destruction mechanisms become far smaller which leads to its fractional contribution to increase despite the absolute contribution being negligible. We only considered the top reactions that contributed to 99 \% of the creation or destruction to limit the number of lines we would have to plot. }
\label{HNC_analysis_output}
\end{figure*}

However, the second reaction does not appear to be the dominant HNC reaction at low temperatures. This can be seen in Figure \ref{HNC_analysis_output} where we plot the fractional contribution of the reactions that are responsible for creating and destroying 99\% of the HNC at each time step alongside the temperature as a function of time. We see that it is in fact the reaction \ce{H3+ + HNC -> HCNH+ + H2} as well as freeze-out responsible for this at low temperatures. As such, we still have an explanation for the two regimes observed, but the oxidisation reaction seems to not play as important a role in our model, suggesting further study might be required. However, the inflection point in \cite{HCN_HNC} is observed to be at 40K, whereas in this work it is at 65K. This can be explained by noting that we consider a wider variety of physical parameter combinations, whereas the other work considered the ones specific to the Orion A Cloud. As such, a quantitative comparison is difficult to make. However, it is reassuring to observe qualitative agreement. Similarly, we observe no real relationship between the cosmic ray ionisation rate and the log-ratio, which is broadly in agreement with the modelling done in \cite{Meijerink2011}. However, there is some disagreement with observations as seen in \cite{Behrens_paper}, though this can be attributed to them considering a larger range of cosmic ray ionisation rates. In that paper, the ratio was found to decrease as the cosmic ray ionisation rate increased, though this was only in the presence of mechanical heating which we do not consider here.

We observe that for metallicity, we have the same ``tailing-off" effect that we have observed previously, though this is only in the marginal case in Figure \ref{HCN_HNC_dependence_plots}. This is the case for metallicity values between 0 and 1. Again, we can attribute this to increased competition for the individual atomic species which results in the ratio decreasing.

\cite{Bayet2012} considered a gas density of $10^{4}$ cm$^{-3}$, radiation field values of 1 Habing, a cosmic ray ionisation rate of 5.0 $\times 10^{-17}$ s$^{-1}$ and metallicities between 1 and 5. For metallicities between 1 and 2, we see a roughly linear marginal increase in the log-ratio. This is in line with what was observed in \cite{Bayet2012} in which an increase in the metallicity results in a linear increase in the log-abundances of HCN and HNC with the HCN having a steeper increase with metallicity. This suggests that their ratio would also increase linearly.

\subsubsection{HCN/CS}

We now consider another tracer, the HCN to CS ratio. This ratio has received significant interest in recent years \citep{Izumi_2013, Izumi_2016, Josh_paper}, with one of the reasons being the fact that both HCN and CS are dense gas tracers \citep{Viti_2017}, with the HCN(4-3)/CS(2-1) ratio being a good tracer of active galactic nuclei (AGN) activity. Just as for the ratio of HCN to HNC, we now wish to obtain a sense of the relationship of the five features of interest with this ratio. 

\begin{figure*}
\includegraphics[width=2.1\columnwidth]{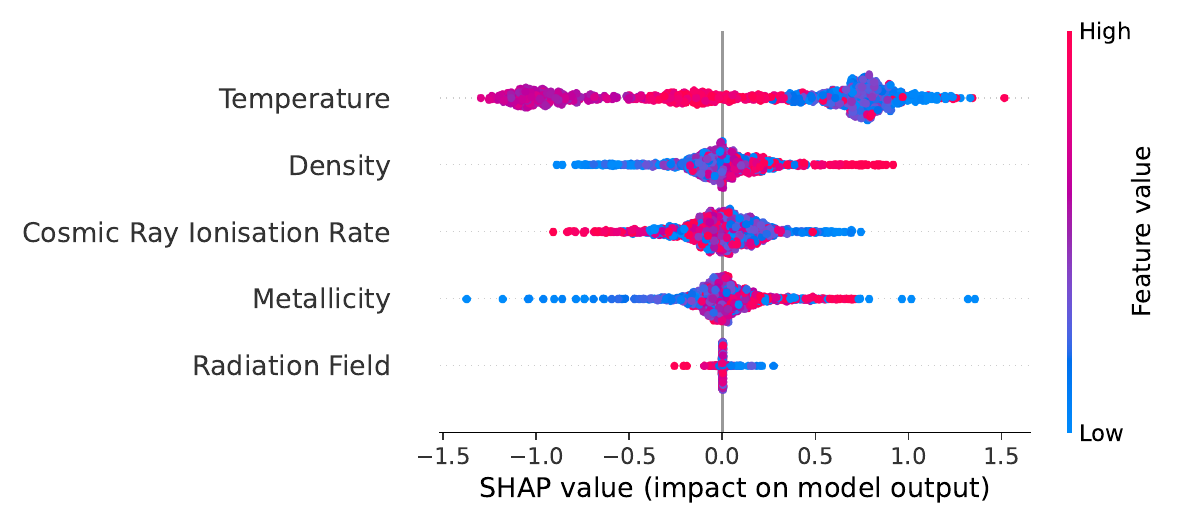}
\caption{A beeswarm plot of the various physical parameters demonstrating their relative importance in predicting the log-ratio of HCN to CS. We observe that temperature has the largest impact on the model output with ($\hat{I}_{T} = 0.55$). Density is also found to have a significant impact ($\hat{I}_{n} = 0.16$), which makes sense as it is seen both HCN and CS are dense gas tracers. The next most important features are cosmic ray ionisation rate ($\hat{I}_{\zeta} = 0.15$), followed by the metallicity ($\hat{I}_{m_z} = 0.13$) and the radiation field ($\hat{I}_{\psi} = 0.01$).}
\label{HCN_CS_dot_plot}
\end{figure*}

We begin by considering the relative importance of the five features. Figure \ref{HCN_CS_dot_plot} is a beeswarm plot demonstrating this. We observe that temperature is once again the most relevant feature followed by density, cosmic ray ionisation rate, metallicity and radiation field.

\begin{figure*}
\includegraphics[width=2.1\columnwidth]{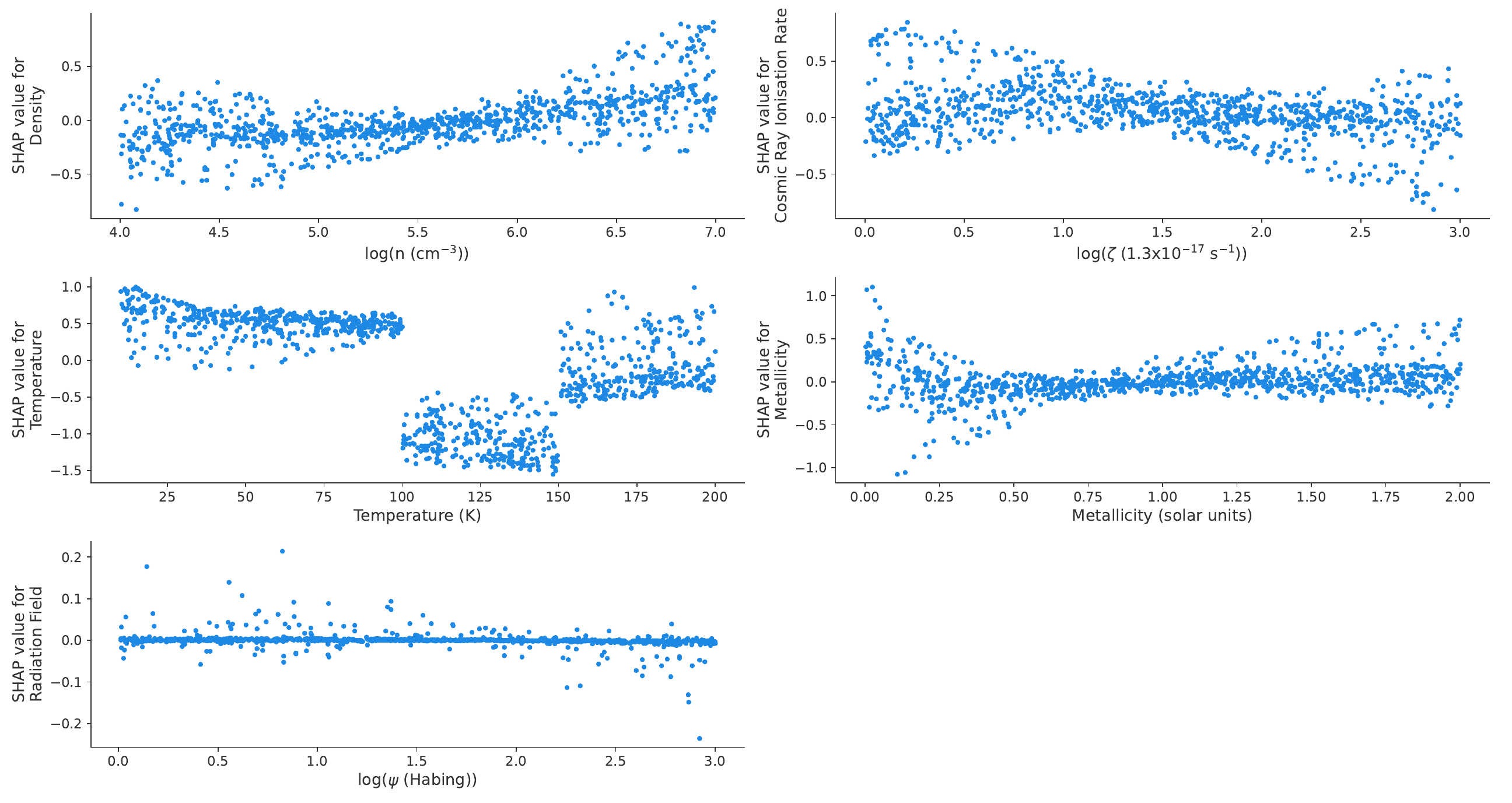}
\caption{A plot of the SHAP values for the various features (besides the radiation field) as a function of the feature values used to predict the log-ratio of HCN to CS. What we observe is that there exist three separate temperature regimes under which the final abundance is relatively constant. We also notice an increase in the SHAP value as the log-density increases.}
\label{HCN_CS_dependence_plots}
\end{figure*}

\begin{figure*}
\includegraphics[width=2.1\columnwidth]{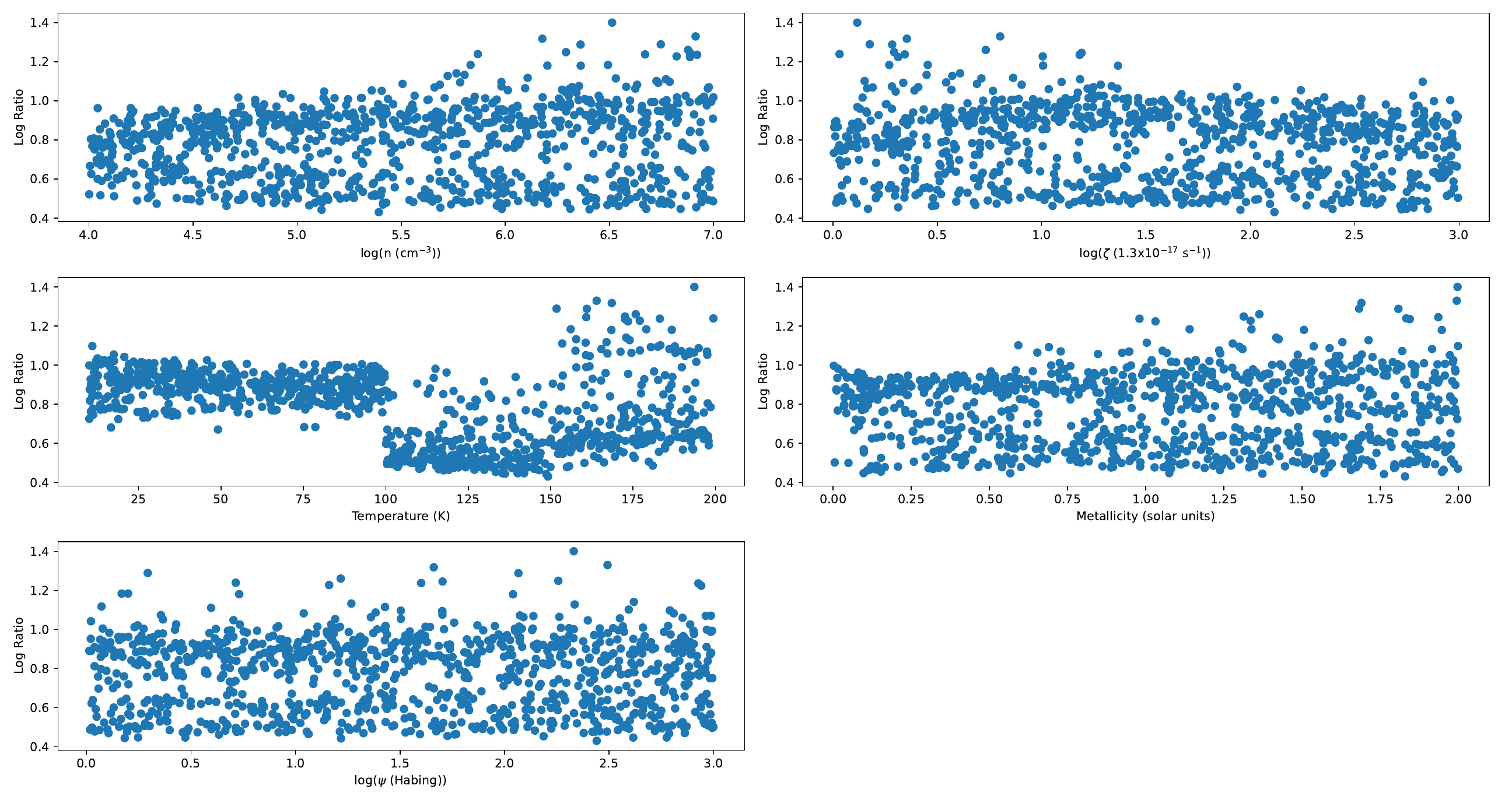}
\caption{A plot of the log-abundance of HCN/CS ratio as a function of the various features. To calculate the log-ratio for a given data point, we sum up the importance values of each feature for that data point. We observe that only temperature maintains a clear trend relative to what we observed in Figure \ref{HCN_CS_dependence_plots}. For the other features, we have no discernible trend which can be attributed to the feature importances nullifying each other.}
\label{HCN_CS_native_plots}
\end{figure*}

We consider this more in Figures \ref{HCN_CS_dependence_plots} and \ref{HCN_CS_native_plots}. There is a clear quasi-linear relationship between the log-ratio and the log-density which supports the idea that the ratio could serve as a density tracer. The cosmic ray ionisation rate and the radiation field do not appear to have discernible relationships with the ratio. We find there is not a monotonic relationship with temperature. In fact, we once again seem to observe three separate temperature regimes. The former shows the SHAP value as a function of the feature value, which means it shows the marginal effect of each feature. The latter considers the abundance as a function of each feature. As we discussed earlier, the abundances plotted are derived from summing the marginal effects of all the features.

\begin{figure*}
\includegraphics[width=2.1\columnwidth]{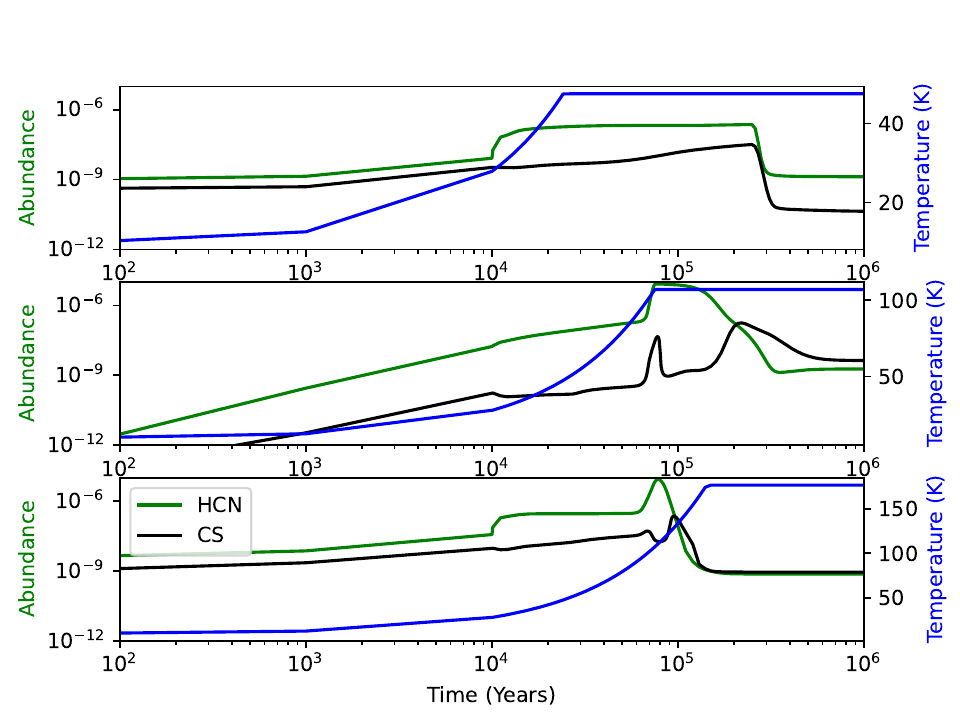}
\caption{A plot of the abundances of HCN and CS as a function of time for three different temperatures taken from the dataset: 47 K, 107 K and 176 K, each of which is within one of the three temperature regimes we observe in the dependence plots for the HCN/CS ratio.}
\label{HCN_CS_abundances}
\end{figure*}

\begin{figure*}
\includegraphics[width=2.1\columnwidth]{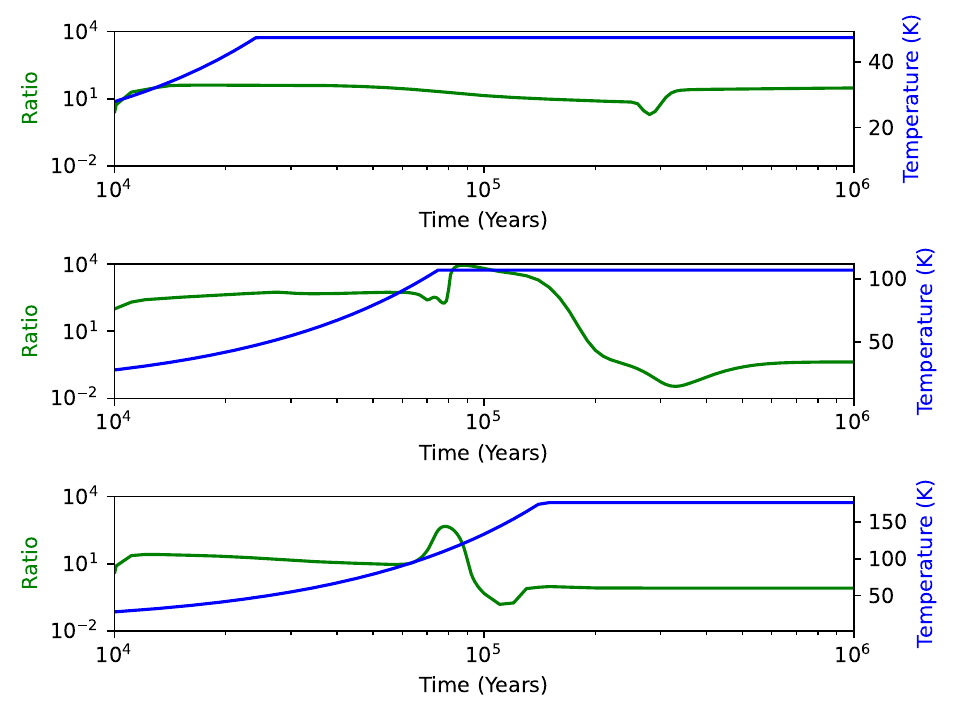}
\caption{A plot of the ratio of HCN to CS as a function of time for three different temperatures taken from the dataset: 47 K, 107 K and 176 K, each of which is within one of the three temperature regimes we observe in the dependence plots for the HCN/CS ratio.}
\label{HCN_CS_ratios}
\end{figure*}

We observe that for the temperature variable there are three separate regimes of interest when it comes to the log-ratio: one for below 100K, one for between 100 and 150 K and another for above 150 K. To start off with, we plot the temporal evolution of the abundances of the two molecules and the temperature in Figure \ref{HCN_CS_abundances} for three different values of the final temperature: 47K, 105K and 176K. These were plotted using UCLCHEM. Note that these temperatures are not special in any way, but they are simply chosen as examples to illustrate the points we wish to discuss. Each of these temperatures falls within one of the three different regimes we observe in Figure \ref{HCN_CS_dependence_plots} and were taken from the dataset. We also plot a time series of the ratio in Figure \ref{HCN_CS_ratios}.

We observe that at 47K, we initially have a large build-up of HCN until about $10^{5}$ years. CS is also built-up, but not to the same extent. After this point, both abundances drop sharply, though the CS drops far more, leading to an increase in the value of the ratio. However, for 107K the abundance of CS exceeds that of HCN leading to a smaller HCN/CS ratio. This is still true for 176K, but CS approaches HCN's abundance much more closely. 

In the low-temperature (<100K) regime, the dominant destruction reaction of HCN is 
\ce{H3+ + HCN -> HCNH+ + H2}. Once the maximum temperature is reached, the main formation reactions are 

\begin{equation}
\ce{HCNH+ + E- -> HCN + H}
\end{equation}

\begin{equation}
\ce{CH + NO -> HCN + O} 
\end{equation}

\begin{equation}
\ce{NH3 + CN -> HCN + NH2}. 
\end{equation}

However, the \ce{NH3}-based reaction becomes less efficient over time at this temperature and is replaced by \ce{N + HCO -> HCN + O}.

In the mid-temperature regime (100K-150K), the major formation routes are:

\begin{equation}
\ce{NH3 + CN -> HCN + NH2}
\end{equation}

\begin{equation}
\ce{NH3 + HCNH+ -> HCN + NH4+}
\end{equation}

\begin{equation}
\ce{CH + NO -> HCN + O}
\end{equation}

\begin{equation}
\ce{HCNH+ + H2CO -> H3CO+ + HCN} 
\end{equation}

\begin{equation}
\ce{N + HCO -> HCN + O}
\end{equation}

\begin{equation}
\ce{CN + HCO -> CO + HCN}
\end{equation}

with the major destruction routes being: 

\begin{equation}
\ce{H3O+ + HCN -> HCNH+ + H2O}
\end{equation}

\begin{equation}
\ce{HCN + CRPHOT -> CN + H}
\end{equation}

\begin{equation}
\ce{H3+ + HCN -> HCNH+ + H2}
\end{equation}

\begin{equation}
\ce{HCN + H3CO+ -> H2CO + HCNH+} 
\end{equation}

\begin{equation}
\ce{CH3+ + HCN -> CH3CNH+ + PHOTON}
\end{equation}

with the final reaction becoming less efficient after about $2.3 \times 10^{5}$ years. 

In the high-temperature regime (>150K), the major HCN reserves are built up until $7.7 \times 10^{4}$ years via these reactions: 

\begin{equation}
\ce{NH3 + CN -> HCN + NH2}
\end{equation}

\begin{equation}
\ce{CH + NO -> HCN + O}
\end{equation}

\begin{equation}
\ce{HCNH+ + H2CO -> H3CO+ + HCN} 
\end{equation}

\begin{equation}
\ce{HCNH+ + E- -> HCN + H} 
\end{equation}

\begin{equation}
\ce{N + HCO -> HCN + O}
\end{equation}

\begin{equation}
\ce{H + H2CN -> HCN + H2}.
\end{equation}

The reactions primarily responsible for the destruction are: 

\begin{equation}
\ce{H3O+ + HCN -> HCNH+ + H2O}
\end{equation}

\begin{equation}
\ce{HCN + CRPHOT -> CN + H}
\end{equation}

\begin{equation}
\ce{H3+ + HCN -> HCNH+ + H2}
\end{equation}

The aforementioned destruction mechanisms are more efficient in the mid-temperature range than in the high-temperature range. This explains why the value of the ratio drops between 100 and 150 K.

We observe a weak linear relationship between the SHAP value and the metallicity. This is in line with what has been observed previously \citep{molecular_metallicity_tracers}. In that work, they considered molecular regions of galaxies with metallicities ranging from 0.1 - 0.6, temperatures between 90 and 220 K with the remainder of conditions not listed in the paper. With the exception of the 200 - 220 K range, the listed conditions overlap with the ones in this work. What they found is that they were able to obtain a separate linear function fitting the log-ratio to the metallicity for each visual extinction value. We know that the greater the visual extinction, the greater the final density of the cloud. Furthermore, fixing the visual extinction and therefore the density fixes the final temperature that our cloud reaches during the warm-up phase. Cosmic ray ionisation rates and the radiation field are also taken to be constant in the observed galaxies. This means that each linear relationship provided in \cite{molecular_metallicity_tracers} gives the relationship between the log-ratio and the metallicity when our other four parameters are fixed. As such, it is sensible to state that there is qualitative agreement between the linear marginal SHAP relationship for metallicity in Figure \ref{HCN_CS_dependence_plots} and the relationships found in \cite{molecular_metallicity_tracers}, as both of these assume the other parameters are fixed. Once again, it makes little sense to compare the exact numbers as we consider a far wider range of conditions. However, the qualitative similarity lends support to the validity of this methodology. The fact that the metallicity going to zero does not cause a tail-off in the value of the ratio suggests that there is another reaction compensating for the depletion of this ratio.

\section{Conclusion}\label{sec:conclusion}
In this work, we present the first application of machine learning interpretability techniques to better understand the effect of various physical parameters on molecular abundances. We trained an XGBoost statistical emulator to replicate the outputs of our chemical model, UCLCHEM. From this, we used SHAP to determine a relative ranking of feature importance as well as to identify the nature of the relationships between the input parameters and the output of interest. A quantitative measure for the relative feature importance was also presented.

This work essentially presents a  sensitivity analysis, but is different in many ways to previous studies. This is the first time that the concept of machine learning interpretability has been applied in astrochemistry to consider the impacts of various parameters on abundances. Our methodology offers a number of advantages. In the first instance, by training a statistical emulator to replace our forward model, UCLCHEM, we are able to significantly reduce the time taken per forward model evaluation, therefore allowing for a much larger grid to be evaluated. Additionally, we are able to quantify the relative importances of the various features as well as comment on the marginal impacts of each of the features.

The main takeaways from this work for the various outputs are as follows: 

\begin{itemize}
    \item \ce{H2O} and CO's gas phase abundances depend strongly on the metallicity, which we relate to the fact that a low metallicity results in the production of each molecule being constrained by the amount of the less abundant atomic element (O and C, respectively). 
    \item \ce{NH3} has a strong temperature dependence. There exist two temperature ranges (< 100K and > 100 K) for which the abundance is constant. We are able to relate this to the chemical reactions in our network and find that the increased temperature results in an increase in the destruction pathways. 
    \item We are able to confirm that the HCN/HNC ratio can serve as a cosmic thermometer and find a two-part linear relationship with temperature as in \cite{HCN_HNC}. However, the dominant HNC destruction at low temperatures is found to be \ce{H3+ + HNC -> HCNH+ + H2} instead of \ce{HNC + O -> NH + CO}. We also find a linear relationship between the metallicity and the log-ratio in the range 1-2 which matches what we find in \cite{Bayet2012}.
    \item For the HCN/CS ratio, we observe that it serves as a density tracer, as expected. Furthermore, we once again observe three separate regimes for the temperature dependence, which we are able to relate to the chemistry. 
\end{itemize}

Another point of interest is that the metallicity parameter often, but not always, leads to a ''tailing-off effect" in the abundance in the limit of the metallicity going to zero. This was the case for \ce{H2O}, CO and the HCN/HNC ratio. However, we did not observe this for \ce{NH3} and the HCN/CS ratio. This suggests that despite the scaling down of the metals, there are other reactions that compensate by creating more of the respective molecule from the limited resources. Further work should consider this in more depth potentially by applying SHAP to a reaction network.

Throughout this work, we have observed similarities between our results and what has been discussed in the literature. This is encouraging. However, it is difficult to make direct quantitative comparisons, as we consider a wide range of physical parameter combinations. On the other hand, the literature we cited considered actual observations. A follow-up study would need to sample the training data for the machine learning model more precisely in order to be able to better model and understand the relationships between inputs and outputs for a specific astronomical object.

\section*{Acknowledgements}
We thank the anonymous referee for their constructive comments that improved the quality of the manuscript. J. Heyl is funded by an STFC studentship in Data-Intensive Science (grant number ST/P006736/1). S. Viti acknowledges support from the European Union’s Horizon 2020 research and innovation programme under the Marie Skłodowska-Curie grant agreement No 811312 for the project ``Astro-Chemical Origins” (ACO). This work was also supported by European Research Council (ERC) Advanced Grant MOPPEX 833460. 

\section*{Data Availability}
The data underlying this article are available in the article and in its online supplementary material.



\bibliographystyle{mnras}
\bibliography{references} 




\appendix
\section{Supplementary Tables}

\begin{table}
\begin{tabular}{ll}
\hline
\textbf{Species} & \textbf{Abundance (relative to H nuclei)} \\ \hline
He               & $1.00 \times 10^{-3}$                                      \\
C                & $1.77 \times 10^{-4}$                     \\
O                & $3.34 \times 10^{-4}$                     \\
N                & $6.18 \times 10^{-5}$                     \\
S                & $3.51 \times 10^{-6}$                     \\
Mg               & $2.25 \times 10^{-6}$                     \\
Si               & $1.78 \times 10^{-6}$                     \\
Cl               & $3.39 \times 10^{-8}$                     \\ 
P                & $7.78 \times 10^{-8}$                     \\
Fe               & $2.01 \times 10^{-7}$                     \\
F                & $3.60 \times 10^{-8}$                     \\ \hline
\end{tabular}
\caption{Initial elemental abundances used in UCLCHEM.}
\label{elemental_abundances}
\end{table}

\begin{table}
\begin{tabular}{llllll}
\hline
\textbf{Species/Ratio} & \textbf{$\hat{I}_{n}$} & \textbf{$\hat{I}_{\zeta}$} & \textbf{$\hat{I}_T$} & \textbf{$\hat{I}_{m_z}$} & \textbf{$\hat{I}_{\psi}$} \\ \hline
\ce{H2O}                    &          0.17     &               0.15   &          0.11     &                 0.54        &            0.02     \\
\ce{CO}                   &        0.01      &           0.04       &       0.04       &       0.91                  &         0.00        \\
\ce{NH3}                    &      0.11         &    0.14              &    0.54           &       0.17                  &       0.03          \\
HCN/HNC                &     0.03          &       0.03           &      0.70         &         0.24                &          0.00       \\
HCN/CS                 &  0.16             &   0.15               &     0.55          &        0.13                 &     0.01           \\ \hline
\end{tabular}
\caption{Table summarising the $\hat{I}_i$ for each parameter $i$.}
\label{relative_importance_table}
\end{table}

\begin{table}
\begin{tabular}{ll}
\hline
\textbf{Species/Ratio} & \textbf{Range of Values}                   \\ \hline
\ce{H2O}                    & $7.7 \times 10^{-12} - 3.8 \times 10^{-6}$ (relative to $n_H$)\\
CO                     & $9.8 \times 10^{-11} - 3.4 \times 10^{-4}$ (relative to $n_H$)\\
\ce{NH3}                    & $1.0 \times 10^{-12} - 1.2 \times 10^{-12}$ (relative to $n_H$) \\
HCN/HNC                & $1.0 - 2418.0$                               \\
HCN/CS                 & $6.0 \times 10^{-3} - 22091.9$                          \\ \hline
\end{tabular}
\caption{Table summarising the range of values of the outputs of the abundances and ratios of interest. In the case of \ce{NH3}, the lower bound of our values has been clipped at $10^{-12}$ as dicussed in the text.}
\label{output_abundances}
\end{table}


\bsp	
\label{lastpage}
\end{document}